\documentclass[a4paper,12pt]{article}
\usepackage{jheppub}

\pdfoutput=1
 \usepackage{hyperref}
\hypersetup{colorlinks=true,linkcolor=magenta,anchorcolor=green,citecolor=cyan,filecolor=black,menucolor=black,urlcolor=brown}
\usepackage{enumerate}
\usepackage{graphicx} 
\usepackage{graphicx}
\usepackage{amsmath,amssymb}

\usepackage[usenames,dvipsnames,table]{xcolor}

\usepackage{tikz}

\usepackage{comment}
\usepackage{graphicx}
\usepackage{subfigure}

\def \be {\begin{equation}}
\def \ee {\end{equation}}

\usepackage{xcolor}   
\usepackage{ulem}    
\newcommand{\add}[1]{#1}
\newcommand{\del}[1]{}

\begin{document}
\preprint{LAPTH-003/25}
\title{HEFT Numerators from Kinematic  Algebra}
\author[a]{Chih-Hao Fu,}
\affiliation[a]{Institute of Mathematics, Henan Academy of Sciences,
NO.228, Chongshi Village, Zhengdong New District, Zhengzhou, Henan 450046, P.R. China}
\author[b]{Pierre~Vanhove,}
\affiliation[b]{Institut de Physique Th\'eorique, Universit\'e  Paris-Saclay, CEA, CNRS, F-91191 Gif-sur-Yvette Cedex, France}
\author[c,d]{and Yihong Wang}
\affiliation[c]{
College of Mathematical Sciences, Harbin Engineering University,
145 Nantong Street, Nangang District, Harbin 150001, P. R. China
}
\affiliation[d]{Laboratoire d'Annecy-le-Vieux de Physique Th\'eorique (LAPTh), CNRS and Universit\'e Savoie Mont-Blanc, Annecy, 74940, France}
\abstract{
{ We  derive the} kinematic numerator factors for heavy-mass effective field theory   from  
\add{ the } field theory limit of the string theory vertex operator kinematic algebra introduced in~\cite{Fu:2018hpu}. The kinematic numerators are 
{ derived as} 
correlators of nested commutators of gluon vertex operators evaluated between massive tachyonic vertex operators. The resulting numerators are given by products of structure constants of the vertex operator algebra which \add{yield}  
gauge invariant expressions.   The computation of the nested commutators leads to a natural organisation in the form of rooted trees, endowed with an order that facilitates the enumeration of the various contributions. This kinematic algebra gives a string theory understanding of  the field theory fusion rules for constructing the  heavy-mass effective field theory numerator of~\cite{Brandhuber:2021kpo,Brandhuber:2021bsf}.
}

\maketitle


\section{Introduction}

\label{sec:introduction}

The colour-kinematics duality~\cite{Bern:2008qj,Bern:2010ue} is a reformulation 
that provides a unified treatment of scattering amplitudes in gauge theory, gravity, and a variety of effective field theories. In this formulation, 
{the dependence on momentum and colour is}
 introduced symmetrically into the amplitude numerators through {the} structure constants. As a consequence, the same formulation describes gravitational amplitudes when the colour algebra is replaced by another copy of the kinematic data. This provides an advantageous alternative for calculating gravity amplitudes (see~\cite{Bern:2019prr,Bern:2022wqg,Carrasco:2015iwa,Adamo:2022dcm} for reviews). In recent years, tremendous progress has been made in understanding colour-kinematic duality through various approaches, especially at the loop level~\cite{Bern:2024vqs}. 
 They include methods from
string theory~\cite{Mafra:2011kj,Mafra:2011nv,Ochirov:2013xba,Tourkine:2016bak,Ochirov:2017jby,Fu:2018hpu,Casali:2020knc,
Stieberger:2022lss,Monteiro:2022lwm,Stieberger:2023nol}, homotopy algebras~\cite{Reiterer:2019dys,Borsten:2020zgj,Borsten:2021hua,Borsten:2021gyl,
Escudero:2022zdz,Borsten:2022ouu,Borsten:2022vtg,Bonezzi:2022yuh,
Borsten:2023reb,Borsten:2023ned,Bonezzi:2022bse,Bonezzi:2023lkx,
Bonezzi:2023ciu,Bonezzi:2024dlv,Bonezzi:2023pox,Szabo:2023cmv}, and  fusion rules~\cite{Chen:2019ywi,Chen:2021chy,Brandhuber:2021kpo,Brandhuber:2021bsf,
Chen:2022nei,Brandhuber:2022enp,Chen:2023ekh,Chen:2024gkj,
Bjerrum-Bohr:2024fbt}. 

The colour-kinematic duality, together 
 {with} 
the double-copy procedure, 
{has found important applications}
 in 
{ evaluating} 
 amplitudes for the post-Minkowskian expansion of gravitationally interacting massive binaries~\cite{Bern:2019nnu,Bern:2019crd,Bern:2022jvn,
Bern:2020buy,Bjerrum-Bohr:2020syg,Bjerrum-Bohr:2021wwt,
Goldberger:2016iau,Shen:2018ebu,Shi:2021qsb,
Bjerrum-Bohr:2023jau,Brandhuber:2023hhy,Bjerrum-Bohr:2023iey,
Brandhuber:2023hhl,Chen:2024mmm,Chen:2024bpf}.
The heavy mass effective field theory (HEFT) tree-level amplitudes are the emission of multi-gluons or multi-gravitons from a massive scalar line, in Yang-Mills and gravity theory respectively. 
One can arrange the (colour-ordered) tree-level multi-gluon emission from a scalar line of momentum $p$  as a sum over cubic graphs 
\begin{equation}\label{e:AmpYM}
  A^{\rm YM}(-p,1,\dots,n,p') = \sum_{i~\textrm{cubic graphs}} {\mathcal N_i\over d_i},  
\end{equation}
such that  the multi-graviton emission from a scalar momentum line $p$ reads 
\begin{equation}
  M^{\rm Gravity}(-p,1,\dots,n,p') = \sum_{i~\textrm{cubic graphs}} {\mathcal N_i^2\over d_i}                  \,.
\end{equation}
A gauge invariant form for the numerator factors has been obtained  using the fusion product between two heavy-mass currents~\cite{Brandhuber:2021kpo,Brandhuber:2021bsf} or as an average  of the colour-ordered amplitude by the momentum kernel~\cite{Bjerrum-Bohr:2010pnr,Mafra:2011nw} and the exponential formalism~\cite{Bjerrum-Bohr:2020syg}. 
The expression for the multi-graviton emission is particularly useful for carrying out the multi-soft graviton expansion and for organizing the post-Minkowskian amplitude in an expansion of the ratio of the masses of the binaries~\cite{Brandhuber:2021eyq,Bjerrum-Bohr:2021wwt}.

The numerators $\mathcal N_i$ in~\eqref{e:AmpYM} are not unique, but it was shown in~\cite{Brandhuber:2021kpo,Brandhuber:2021bsf} that they can be chosen to have particularly nice properties:
(1) they depend only on the field strength
$F^i_{\mu\nu}=\epsilon^i_{[\mu} k^i_{\nu]}$ of the gluons, (2) \del
{the expression of } the numerator {factors are } 
naturally organised in terms of nested commutators, reflecting the
tree { amplitude} structure of gluon or graviton emission{s}. These properties have been linked to the fusion product between two heavy-mass currents~\cite{Brandhuber:2021kpo} and suggest  an underlying quasi-Hopf algebra~\cite{Brandhuber:2021kpo,Brandhuber:2021bsf}.

\medskip
In this paper, we use the vertex operator construction of the
kinematic algebra of~\cite{Fu:2018hpu}, to provide an algorithmic
construction of the gauge invariant numerator factors $\mathcal N_i$.
The derivation is based on the evaluation 
{of} 
the expectation value  between
massive external states of
nested commutators of vertex operators for gluon states. 
 In section~\ref{sec:kinematicalgebra}, we define the kinematical
 algebra of the $\alpha'$-weighted commutator between two vertex
 operators. In section~\ref{sec:heft-algebra}, we show that the
 numerator factors $\mathcal N_i$ in~\eqref{e:AmpYM} are given by the
 field theory limit of the expectation value of nested commutators of
 gluon vertex operators betwee{n} two tachyonic vertex operators
 associated  {with} the massive scalar particle.
In section~\ref{sec:fermion}, we describe how to extend this
construction to massive fermionic states. Other massive external
states could be considered along the same formalism but will be
discussed here.
Section~\ref{sec:numerator-derivation} describes an algorithmic
derivation of the numerator factors for the massive scalar case. The
evaluation of the nested commutators between the vertex operators  is
naturally organised into a rooted tree equipped with an order of the
legs. We explain  that the nested commutators are naturally evaluated
using the Shapovalov form formalism {of}~\cite{Fu:2022esi}. The numerator factors are not unique, but we explain in section~\ref{sec:manifest}, how a manifest gauge invariant form is obtained. In the appendix~\ref{sec:matchQMUL}, we give details on the three and four gluons emission numerators and show how to match the construction of~\cite{Brandhuber:2021kpo, Brandhuber:2021bsf} term by term. We conclude in section~\ref{sec:conclusion}.

The algorithmic construction has been implemented in a \texttt{Mathematica} code available in the GitHub repository~\href{https://github.com/Yi-hongWang/Stringy-Numerator}{https://github.com/Yi-hongWang/Stringy-Numerators}. 


\section{The kinematic algebra and structure constants}
\label{sec:kinematicalgebra}

We start with the $\alpha'$-deformed commutator between two vertex operators~\cite{Ma:2011um,Carrasco:2016ygv}
\begin{equation}\label{e:commutator}
[V_{1},V_{2}]_{\alpha'}=V_{1}V_{2}-e^{-i\pi\alpha'k_{1}\cdot k_{2}}V_{2}V_{1}.
\end{equation}
We will be working with unintegrated vertex operators for the massive tachyonic external line in the $-1$-ghost picture
\begin{equation}\label{e:vopghost}
    V^{\textrm{scalar}}(p)=:c(t) e^{i p\cdot X(t)}: ,
\end{equation}
and the integrated gluon vertex operators in the 0-ghost picture
\begin{equation}
  V^{\textrm{vector}}(\epsilon,k):=\int_{0}^{1} dt  :i \epsilon\cdot \partial_t X(t)   e^{ik\cdot X(t)}:,
\end{equation}
and the $-1$-picture gluon vertex operator
\begin{equation}
    \tilde V^{\textrm{vector}}(\epsilon,k):=:c(t) i\epsilon\cdot \partial_t X(t)   e^{ik\cdot X(t)}:.
\end{equation}
{Additionally,}
we introduce a notation for the generators of kinematic algebra  with the polarisation tensor removed  for vectors
\begin{equation}\label{e:VtoE}
V^{\textrm{vector}}(\epsilon,k)=: i\epsilon_\mu E^{\textrm{vector},\mu}(k),
\end{equation}
and its generalization for  higher rank tensors, which we will meet below.

In  operator representation 
{theory},
the commutator in~\eqref{e:commutator}
is computed by considering the action of the vertex operators together
with the integration as the screening operators in  conformal field
theory.  This  is 
{recognised as}
 a representation of the $q$-deformed Lie
algebra~\cite{Fu:2020frx}, where the momenta are identified as root
vectors, and the deformation parameter is given by
$q=e^{-i\alpha'\pi}$. The generator of this algebra is discussed in
section~\ref{sec:shapovalov}, where the Shapovalov form is used to
simplify the integration rules over the position of the vertex
operators.

\begin{figure}[h!] 
\centering 
\subfigure[]{ \includegraphics[width=3.5cm]{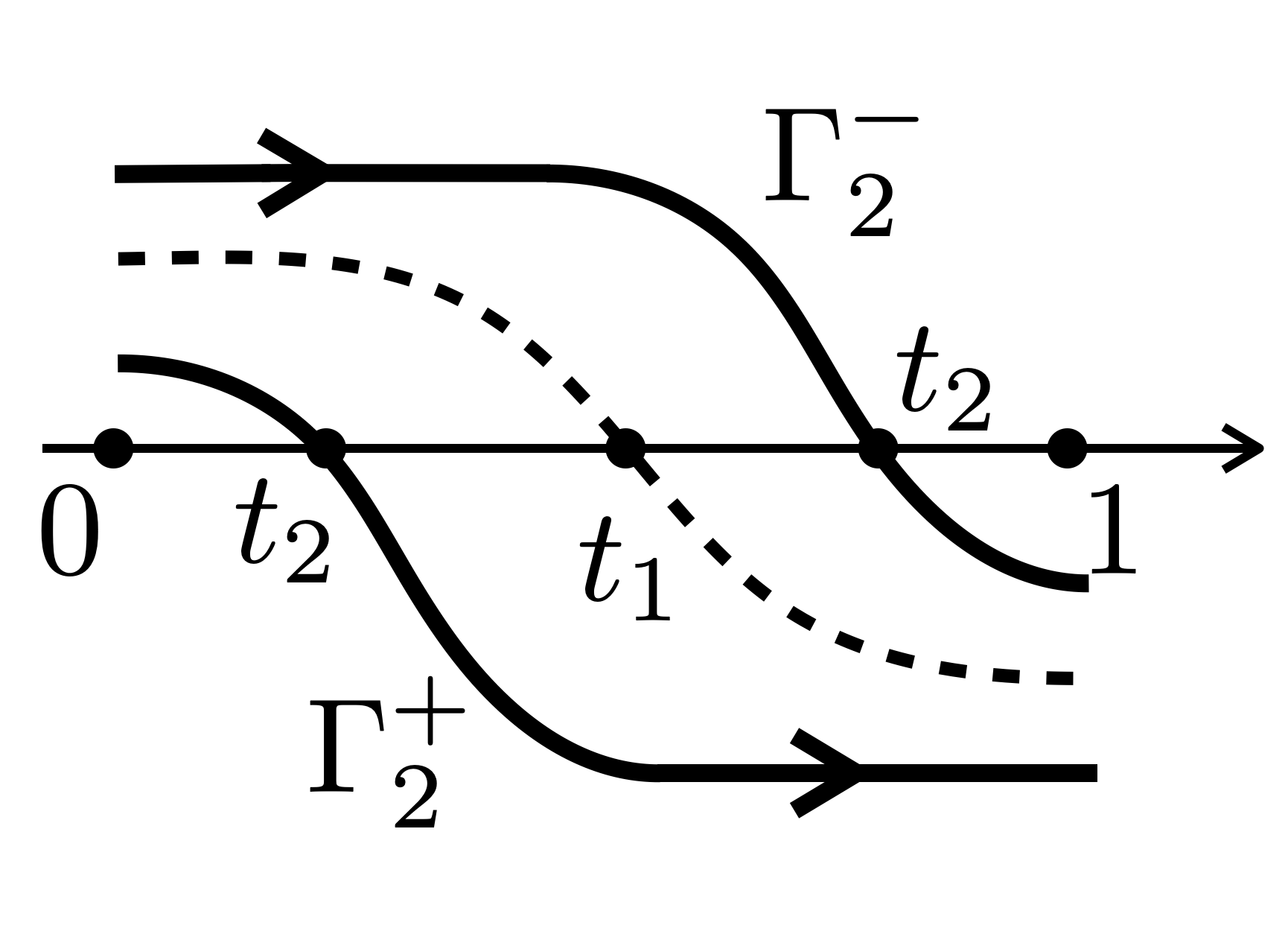}  \label{fig:lines}   }
\includegraphics[width=1cm]{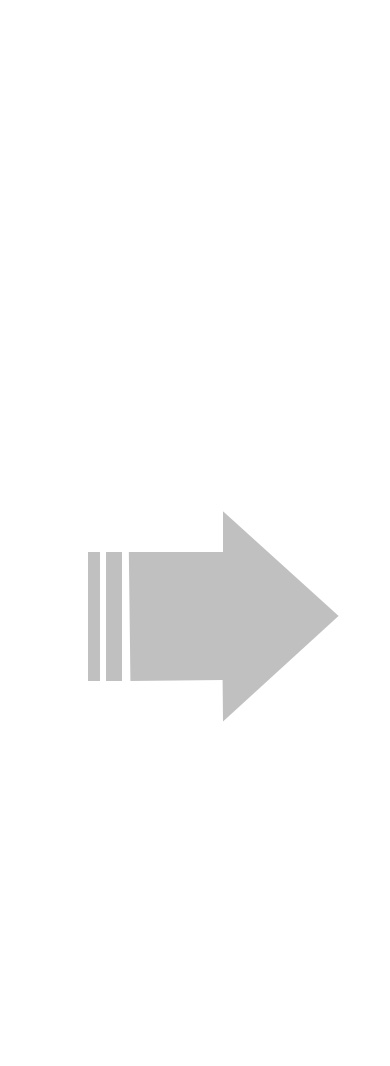} 
\subfigure[]{ \includegraphics[width=3.5cm]{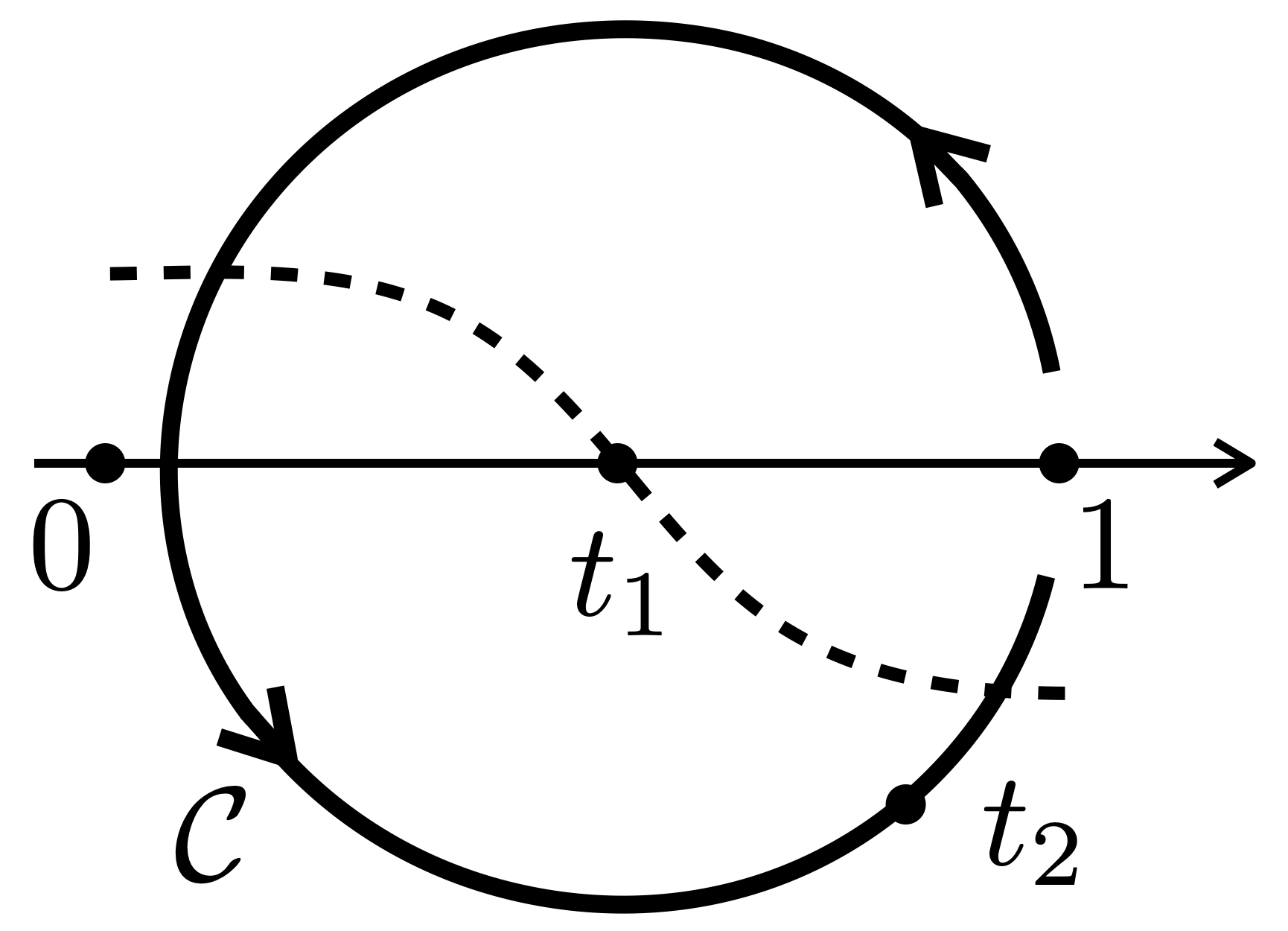}  \label{fig:circle}   }
\caption{Integration contour of a screening operator.  }
\label{fig:contours}
\end{figure}

\medskip
We first explain how to calculate the commutator between different states,
and then we derive the kinematic structure constants. 
In the following sections, we will only need to consider  commutators between integrated vertex operators. Therefore, we will concentrate on this case.
When computing the commutators, we follow the convention introduced
in~\cite{Bjerrum-Bohr:2010pnr}{,  which is} to {position at a
  small distance $i\delta$ on the complex plane}  the vertex operator on the left in an operator product, and then continue analytically. The deformation parameter $q=e^{-i\alpha'\pi}$ allows us to 
{flip  the Koba-Nielsen factor and combine the two terms in the commutator,}
and the result is an integration of one vertex operator over a contour
$\mathcal C$ circling the other, {as represented in Fig.}~\ref{fig:contours}. 
For example, the commutator of two scalars is
\begin{eqnarray}
     [V_{1}^{{\rm scalar}}(k_{1}),V_{2}^{{\rm scalar}}(k_{2})]_{\alpha'} 
     & = &	
     \int_{0}^{1}dt_{1}\int_{\Gamma_{2}^{+}}dt_{2}(t_{1}-t_{2})^{\alpha'k_{1}\cdot k_{2}}:e^{ik_{1}\cdot X(t_{1})}e^{ik_{2}\cdot X(t_{2})}: \label{eq:commutator} \\
     & & -e^{-i\pi\alpha'k_{1}\cdot k_{2}}\int_{\Gamma_{2}^{-}}dt_{2}\int_{0}^{1}dt_{1}(t_{2}-t_{1})^{\alpha'k_{1}\cdot k_{2}}:e^{ik_{1}\cdot X(t_{1})}e^{ik_{2}\cdot X(t_{2})}: \nonumber \\
     & = &	\int_{0}^{1}dt_{1}\int_{\mathcal{C}}dt_{2}(t_{1}-t_{2})^{\alpha'k_{1}\cdot k_{2}}:e^{ik_{1}\cdot X(t_{1})}e^{ik_{2}\cdot X(t_{2})}:.\nonumber
\end{eqnarray}

When the commutator is inserted into an amplitude expression, branch cuts occur, so care must be taken when deforming the contour of integration. 
In this section, to better illustrate the properties of kinematic
algebra, we will 
{ consider first} 
two examples where all momenta are
restricted to the integer lattice $\alpha' k_{i}\cdot
k_{j}\in\mathbb{Z}$ so that we can close the contour in the absence of
branch cuts, and 
{then} 
a final example where we relax this condition.

\begin{enumerate}[(i)]
\item For  two scalar vertex operators $V^{\textrm{scalar}}(k_{1})$, $V^{\textrm{scalar}}(k_{2})$, the commutator
 is zero unless $\alpha'k_{1}\cdot k_{2}$ is negative.  {The}
first non-trivial example is given by states 
{whose} 
lattice momenta satisfy the condition $\alpha'k_{1}\cdot k_{2}=-2$.
In this case equation~(\ref{eq:commutator}) becomes
\begin{equation}
[V^{\rm scalar}_1(k_1),V^{\rm scalar}_2(k_2)]_{\alpha'}= 
\int_{0}^{1} dt_{1} \oint dt_{2}
\frac{1}{(t_{1}-t_{2})^{2}}:e^{ik_{1}\cdot X(t_{1})}e^{ik_{2}\cdot
  X(t_{2})}:,
\end{equation}
because $\alpha'k_{1}\cdot k_{2}$ is a negative integer we can deform
the contour of integration $\mathcal C$ without meeting branch cuts, and the operator product expansion (OPE) of the vertices can be simply
computed via {the} residue theorem, to yield
\begin{equation}
[V^{\rm scalar}(k_1),V^{\rm scalar}(k_2)]_{\alpha'}=\frac{2\pi i}{\sqrt{\alpha'}} \int_{0}^{1} dt_1
:i\sqrt{\alpha'} k_2\cdot X'(t_1)\, e^{i(k_1+k_2)\cdot X(t_1)}:,
\end{equation}
where we have set $X'(t)=\partial_t X(t)$.
So that
\begin{equation}
[V^{\rm scalar}(k_1),V^{\rm scalar}(k_2)]_{\alpha'}= V^{\rm vector}(\sqrt{\alpha'}k_2,k_1+k_2).\label{eq:integer-example-1} 
\end{equation}
The result is the vertex operator of a vector state with 
polarisation tensor $\epsilon=\sqrt{\alpha'}k_{2}$ and momentum $k=k_{1}+k_{2}$, which
satisfy the gauge and mass shell conditions
automatically~\cite{GoddardOlive}.
Using the notation introduced in~\eqref{e:VtoE} we can write this commutation relation as
\begin{equation}
[E^{\text{scalar}}(k_{1}),E^{\text{scalar}}(k_{2})]_{\alpha'}=2\pi i\,
k_{2,\mu}E^{\text{vector},\mu}(k_1+k_2),
\end{equation}
so that with respect to this basis,
the structure constant is $ 2\pi i k_{2,\mu}$. 

\item We consider the commutator between the vertex operator of a scalar and a vector state such that
$\alpha'\,k_{1}\cdot k_{2}=-m$, where $m$ is a large positive integer.
\begin{eqnarray}
 && [V^{\textrm{scalar}}(k_1),V^{\textrm{vector}}(\epsilon_2,k_2)] \\
 && =
 \int_{0}^{1} dt_{1} \oint dt_{2}
:\left(\frac{1}{(t_{1}-t_{2})^{m}}i\epsilon_{2}\cdot
  X'(t_{2})+\frac{-1}{(t_{1}-t_{2})^{m+1}}k_{1}\cdot\epsilon_{2}\right)
e^{ik_{1}\cdot X(t_{1})}e^{ik_{2}\cdot X(t_{2})}:.
\nonumber
\end{eqnarray}
{The} residue theorem then suggests that the two terms can be computed by
taking derivatives $2\pi i/(m-1)!\,\partial_{t_{2}}^{m-1}$ and $2\pi i/m!\,\partial_{t_{2}}^{m}$
before taking the limit $t_{2}\rightarrow t_{1}$. Suppose if we let the one
extra copy of derivative act, bringing down an $ik_{2}\cdot X'(t_{2})$
from the exponential, the two terms can then be combined with a common
denominator. 
{If} we leave the operator as it is, without taking
the remaining derivatives explicitly, the result is a vector-like
vertex 
{operator}
manifestly in gauge invariant form
\begin{eqnarray}
 && [V^{\textrm{scalar}}(k_1),V^{\textrm{vector}}(\epsilon_2,k_2)] 
 \label{eq:integer-example-2} \\
 && =-\frac{2\pi i}{m!}
 \int_{0}^{1}dt_1
  :\partial_{t_{2}}^{m-1}\left(\alpha'\,ik_{1}\cdot F_{2}\cdot X'(t_{2})e^{ik_{1}\cdot X(t_{1})}e^{ik_{2}\cdot X(t_{2})}\right):\Bigr|_{t_{2}\rightarrow t_{1}}.
  \nonumber
\end{eqnarray}
{If we  evaluate the derivatives explicitly, because}
$m$ is a positive integer, 
{the integrand}   {be} a sum of products of
derivatives  $\{X'$,
$X''$, $X'X'$, $\dots$\}.  {We conclude that} the right-hand-side
of~\eqref{eq:integer-example-2}  contains a sum of  contributions
to vectors and higher{-}rank tensors.

\item 
{As a third example,} 
we consider 
the case of the commutator of two vertex operators for vector
states. We relax the condition that the momenta are on an integer
lattice, which is what we need  {for} the heavy mass effective field
theory computation of section~\ref{sec:numerator-derivation}. {The
  commutator evaluates as}
\begin{eqnarray}
&& [V^{\rm vector}(\epsilon_1,k_1),V^{\rm vector}(\epsilon_2,k_2)]  \label{e:V1vecV2vec} \\
&&
 = \int_{0}^{1}dt_1 \int_{\mathcal C } dt_{2}
\,(t_1-t_{2})^{\alpha' k_{1}\cdot k_{2}}:e^{ik_1\cdot X(t_1)}e^{ik_{2}\cdot X(t_{2})} \nonumber\\ 
&& \hspace{0.5cm} \times\left(
  \underset{(1)}{\underbrace{\frac{{\alpha'\epsilon}_{1}\cdot\epsilon_{2}}{(t_1-t_{2})^{2}}}}\,+\,\underset{(2)}{\underbrace{\frac{{\alpha'\epsilon}_{1}\cdot k_{2}}{t_1-t_{2}}(-1)\frac{\alpha'\epsilon_{2}\cdot k_{1}}{t_1-t_{2}}}}+\,\underset{(3)}{\underbrace{\frac{{\alpha'\epsilon}_{1}\cdot k_{2}}{t_1-t_{2}}\,i\epsilon_{2}\cdot X'(t_{2})}} 
  \right.
  \cr
  &&
  \hspace{1.5cm} \left. +\,\underset{(4)}{\underbrace{(-1)\frac{\alpha'\epsilon_{2}\cdot k_{1}}{t_1-t_{2}}\,i{\epsilon}_{1}\cdot X'(t_1)}}+\,\underset{(5)}{\underbrace{i{\epsilon}_{1}\cdot X'(t_1)\,i\epsilon_{2}\cdot X'(t_{2})}}: \right).
  \nonumber
\end{eqnarray}
We can recast all the newly introduced $\epsilon_{2}$ dependent into
the field-strength
$F_{2}^{\mu\nu}=\epsilon_2^{\mu}k_2^{\nu}-\epsilon_2^\nu k_2^\mu$. 
By integrating by parts (IBP) the double pole term  in $(1)$, it combines
with $(2)$ and $(3)$; and we integrate by part  $(4)$ which combines with $(5)$. Using the notation~\eqref{e:VtoE} we obtain at leading order in $\alpha'$,
\begin{equation}
  [{\epsilon}_{1\,\mu}E^{\text{vector},\mu}(k_1),\epsilon_{2\,\nu}E^{\text{vector},\nu}(k_2)]_{\alpha'}
  = f^{[1]}{}_{\mu}E^{\text{vector},\mu}(k_{1}+k_{2})+ f^{[2]}{}_{\mu\nu}E^{\text{tensor},\rho\eta}(k_{1}+k_{2}),\label{eq:4pt-commutator}
\end{equation}
with the vector generator 
{being}
\begin{equation}
 E^{\text{vector},\mu}(k_1+k_2):=
 \int_{0}^{1}dt_1 \int_{\mathcal C } dt_{2}
 \frac{1}{t_{1}-t_{2}}(t_1-t_{2})^{\alpha' k_{1}\cdot
   k_{2}}:iX'^{\mu}(t_{2})e^{i(k_{1}+k_{2})\cdot
   X(t_{2})}:,\label{eq:4pt-vector}
 \end{equation}
and 
the tensor generator
{is given by}
 \begin{equation}
E^{\text{tensor},\mu\nu}(k_{1}+k_{2}):= 
\int_{0}^{1}dt_1 \int_{\mathcal C } dt_{2}
(t_1-t_{2})^{\alpha' k_{1}\cdot k_{2}}:iX'^{\mu}(t_{1})iX'^{\nu}(t_{2})e^{i(k_{1}+k_{2})\cdot X(t_{2})}:,\label{eq:4pt-tensor}
\end{equation}
{whereas the} 
structure constants
{are}
\begin{equation}\label{e:fcste}
  f^{[1]}{}_{\mu}=-\alpha'(\epsilon_{1}\cdot F_{2})_{\mu},\qquad 
    f^{[2]}{}_{\mu\nu}={\epsilon_{1\,\rho}(k_{1}\cdot
    F_{2})_{\eta}\over k_{1}\cdot k_{2}}.
\end{equation}
\end{enumerate}
We have kept the terms that will contribute to the leading order
$\alpha'\to0$ limit, which is enough for the field theory numerator
factors. For higher derivative corrections to the numerator factors as
considered in~\cite{Chen:2024gkj},  one would need to keep 
{contributions of} higher order in $\alpha'$. 
This process generalises iteratively to the computation of nested commutators.
In particular, for the case of vertex operators for vector states with polarisation tensor $\epsilon_{i}$, after integration by parts of the modulus $t_i$, the structure constants will be multiplied by their 
fieldstrength. The result is {generically given by} a sum of vector
and tensor-like vertex operators, with the new $F_{i}$ lining up behind
the existing ones.

The result of nested commutators is  a sum of tensor vertex operators multiplied by associated structure constants. One could wonder if these higher{-}order structure constants are needed or are redundant, 
{in the sense that we may be tempted to} 
combine the right-handside of~\eqref{eq:4pt-commutator} into a single term using a total derivative of the integrand  as in~(\ref{eq:integer-example-2}). But this is 
{actually}
not possible because the result is expanded on two different physical states.


\section{The HEFT numerators from the kinematic algebra}
\label{sec:heft-algebra}

The amplitude in~\eqref{e:AmpYM} is the emission of $n$ gluon of
momenta $k_i$  and polarisation tensor $\epsilon_i$ emitted by a
pair of massive particles of momenta, with momentum conservation and on-shell conditions
\begin{equation}
k_1+\cdots +k_n=p'-p, \qquad k_i^2=0,\qquad p^2=(p')^2,
\end{equation}
which implies that
\begin{equation}
p\cdot \left(k_1+\cdots +k_n\right)=p'\cdot \left(k_1+\cdots +k_n\right)=0.
\end{equation}
 The kinematic numerator factors of the colour-ordered Yang-Mills amplitude can be expressed as the average of the $n-1$ permutations of the  colour-ordered amplitude weighted by the momentum kernel~\cite{Bjerrum-Bohr:2010pnr,Mafra:2011nw}
 \begin{eqnarray}
 && \mathcal N(1,\sigma(2),\dots,\sigma(n)) 
 \label{e:Nstring}\\
 && =\frac{\alpha'^{-n}}{n}\sum_{\rho\in \mathfrak S_{n-1}}\mathcal{S}[\{1,\sigma(2),\dots,\sigma(n)\}^{T}|\rho]|_{p}\,A^{\rm YM}(1,\sigma(2),\dots,\sigma(n),p',-p),
 \nonumber
 \end{eqnarray}
 where $\mathcal{S}[\{1,\sigma(2),\dots,\sigma(n)\}^{T}|\rho]|_{p}$ is the field theory momentum kernel~\cite{Carrasco:2016ldy} with  
reference momentum $p$. We recall that the momentum kernel formalism applies  {to} massive
{states} 
as well, because it only depends on the scalar product of the momenta of the external particles and momentum conservation.

\medskip
 Following the steps of~\cite{Fu:2018hpu}, we see that the numerator factor is given by the expectation value of the commutator of the gluon vertex operators between the  massive tachyon external state of momentum $p$ and $p'$
 \begin{multline}
 \mathcal N(1,\sigma(2),\dots,\sigma(n)) 
 \label{eq:expectation-value} \cr
 =\frac{1}{n} \left(\frac{-i}{ \pi } \right)^{n-1} 
 \lim_{\alpha'\to0} \alpha'^{-2n+1}
 \Bigl\langle p\Bigr|\left[\left[\left[V^{\textrm{vector}}_{1},V^{\textrm{vector}}_{\sigma(2)}\right]_{\alpha'},V^{\textrm{vector}}_{\sigma(3)}\right]_{\alpha'},\dots,V^{\textrm{vector}}_{\sigma(n)}\right]_{\alpha'}\Bigl|p' \Bigr\rangle.
 \end{multline} 
In this expression, the vertex operators for the massive tachyonic scalar are fixed to the position $t_{n+1}=1$ and $t_{n+2}=\infty$. The gluon vertex operator $V_1$ is fixed at position $t_1=0$. For this gluon, we use the shifted gauge invariant polarisation 
\begin{equation}\label{e:barepsilon}
  \bar{\epsilon}_{1}^\mu=\epsilon_{1}^\mu-\frac{p\cdot\epsilon_{1}}{p\cdot
    k_{1}}k_{1}^\mu={(p\cdot F_{1})^{\mu}\over p\cdot k_{1}},
  \end{equation}
 so that it is orthogonal to the scalar momentum $p\cdot \bar \epsilon_1=0$.

\medskip

We calculate the numerator factor $\mathcal N(1,2)$ for the emission of two gluons from the massive scalar line. 
The expression we need to evaluate is given by 
\begin{equation}
    \mathcal N(1,2)=\frac{1}{2} 
     \left(\frac{-i}{ \pi} \right)
    \lim_{\alpha'\to0} \alpha'^{-3}   
    \Bigl\langle p\Bigr|\left[V^{\textrm{vector}}_{1},V^{\textrm{vector}}_{2}\right]_{\alpha'}\Bigl|p' \Bigr\rangle.
\end{equation}
Applying the result of~\eqref{eq:4pt-commutator}, with the gluon $1$  
{carring}
the gauge invariant polarisation tensor $\bar\epsilon_1$ in~\eqref{e:barepsilon} 
{, we get}
\begin{multline}
      \mathcal N(1,2)=-\frac{1}{2} \left(\frac{-i}{ \pi} \right)
      \lim_{\alpha'\to0}  \alpha'^{-2}
      {(p\cdot F_1\cdot F_2)_\mu\over p\cdot k_1} \Bigl\langle p\Bigr|E^{\textrm{vector}\,\mu}(k_1+k_2)\Bigl|p' \Bigr\rangle\cr
+ \frac{1}{2} \left(\frac{-i}{ \pi} \right)
\lim_{\alpha'\to0} \alpha'^{-2}
{(p\cdot F_1)_\mu (k_1\cdot F_2)_\nu\over k_1\cdot k_2} \Bigl\langle p\Bigr|E^{\textrm{tensor}\,\mu\nu}(k_1+k_2)\Bigl|p'\Bigr\rangle  
\label{eq:3pt-numerator-1}   .
\end{multline}
{Note that the} 
contribution of the rank-2 tensor in the second line vanishes. 
{This is }
because $|p\rangle$ is a scalar tachyonic state,
{and} 
the only surviving contribution is proportional to 
\begin{equation}
    p^\mu p^\nu \Bigl\langle p\Bigr| \int_{\mathcal C} dt_{2}(-t_{2})^{\alpha' k_{1}\cdot k_{2}}:e^{ip\cdot X(1)}\,e^{ik_{1}\cdot X(0)}e^{ik_{2}\cdot X(t_{2})}:\Bigl|p' \Bigr\rangle,
\end{equation}
which 
{vanishes} 
when 
{contracting} 
with ${(p\cdot F_1)_\mu (k_1\cdot F_2)_\nu\over k_1\cdot k_2}$ by antisymmetry of the field strength $F_1$.
The vanishing  of the rank-2 tensor contribution is due to our choice
of reference momentum for the polarisation $\bar\epsilon_1$ of gluon
1. {We remark that the higher rank tensor contributes to the
  numerator factor for }   the case of the emission of gluons from a massive fermion line discussed in section~\ref{sec:fermion}.

{In equation (\ref{eq:3pt-numerator-1}), }
we are left with the vector contribution, which evaluates to
\begin{equation}
 \mathcal N(1,2)=i \, {p\cdot F_{1}\cdot F_{2}\cdot p\over 2p\cdot k_{1}}.
  \end{equation}
This reproduces the numerator factor given in~\cite{Brandhuber:2021bsf}.

\medskip
{At higher points,}
the commutator of $n$ vertex operators leads to the sum 
\begin{equation}
\left[\left[\left[V^{\textrm{vector}}_{1},V^{\textrm{vector}}_{\sigma(2)}\right]_{\alpha'},V^{\textrm{vector}}_{\sigma(3)}\right]_{\alpha'},\dots,V^{\textrm{vector}}_{\sigma(n)}\right]_{\alpha'}= \sum_{r=1}^n f^{[r]}_{\mu_1,\dots,\mu_r} \, E^{\textrm{tensor},\mu_1,\dots,\mu_r}(0).
\end{equation}
With the gauge choice in~\eqref{e:barepsilon} for the polarisation of
the gluon at position $t_1=0$, the rank $n$ tensor drops out,  so that
the numerator factor~\eqref{e:Nstring} is given by
\begin{eqnarray}
&& \mathcal N(1,\sigma(2),\dots,\sigma(n)) 
\\
&&=\frac{1}{n} \left(\frac{-i}{ \pi } \right)^{n-1}
\lim_{\alpha'\to0}
\alpha'^{-2n+1}
\Bigl\langle p\Bigr|\left[\left[\left[V^{\textrm{vector}}_{1},V^{\textrm{vector}}_{\sigma(2)}\right]_{\alpha'},V^{\textrm{vector}}_{\sigma(3)}\right]_{\alpha'},\dots,V^{\textrm{vector}}_{\sigma(n)}\right]_{\alpha'}\Bigl|p' \Bigr\rangle 
\nonumber \\
&& = \frac{1}{n} \left(\frac{-i}{ \pi } \right)^{n-1}
\lim_{\alpha'\to0} \alpha'^{-2n+1}
\, \sum_{r=1}^{n} f^{[r]}_{\mu_1,\dots,\mu_r} \, \Bigl\langle p\Bigr|E^{\textrm{tensor},\mu_1,\dots,\mu_r}(0)\Bigl|p' \Bigr\rangle.
\nonumber
\end{eqnarray}
where $\sigma$ is a permutation of the gluon lines and 
\begin{equation}
 E^{\textrm{tensor},\mu_1,\dots,\mu_r}(0)= \int_{\mathcal C} dt_2\cdots \int_{\mathcal C} dt_n  \prod_{1\leq i,j\leq n} (t_i-t_j)^{\alpha' k_i\cdot k_j} :\mathcal X^{\mu_1,\dots,\mu_r}(t_1,\dots,t_n)\prod_{i=1}^n e^{i k_i X(t_i)}:,
\end{equation}
where $t_1=0$ and the rank $r$ tensor $\mathcal
X^{\mu_1,\dots,\mu_r}(t_1,\dots,t_n)$ arises from the contraction for
the $X'(t)$ and the plane wave {factors}.
The double poles are removed by integration by parts, 
as in the example in eq.~\eqref{e:V1vecV2vec}. 
{To carry out the remaining calculation explicitly, we may choose to integrate }
by parts to bring $\mathcal X^{\mu_1,\dots,\mu_r}(t_1,\dots,t_n)$ into a sum of Parke-Taylor
factors with tensorial coefficients depending on $X'^{\mu_i}(t_i)$ and $\eta^{\mu_j\mu_k}$, 
which have a simple $\alpha'\rightarrow0$ 
limit~\cite{Mafra:2011nw,Carrasco:2016ldy,Mafra:2016mcc,Carrasco:2016ygv},
or  
{we may choose to compute the integrals directly}
using the integration rules
of~\cite{Baadsgaard:2015voa,Lam:2015sqb}. For external states, given by massive (tachyonic) scalar, the highest rank $n$ tensor does not contribute to the expectation value, but this is not the case for general massive external states.


\section{HEFT Numerators for external massive fermions states}\label{sec:fermion}

The numerator factors for the emission of $n$-gluons from  a pair of massive fermions  are 
now given by the average over 
the  colour-ordered Yang-Mills multi-gluon emission  
{amplitudes} from two massive fermions 
{weighted by momentum kernel. }
\begin{multline}
  \mathcal N^{\rm fermions}(1,\sigma(2),\dots,\sigma(n)) 
 \label{e:NstringFermions}\cr
  =\frac{\alpha'^{-n}}{n}\sum_{\rho\in \mathfrak S_{n-1}}\mathcal{S}[\{1,\sigma(2),\dots,\sigma(n)\}^{T}|\rho]|_{p}\,A^{\rm YM}(1,\sigma(2),\dots,\sigma(n),v(-p'),u(p)),
  \end{multline}
where the fermion state satisfy the on-shell condition $(\gamma^\mu p_\mu+m)u(p)=0$ and $v(-p')(-\gamma^\mu p'_\mu-m)=0$.
 
They are computed using the same procedure as before, the only change is the last step involving the evaluation of the expectation value, which is  now done between fermionic massive external states. We work with the light-cone gauge vertex operators for the gluon and fermion given in section~7.4 of~\cite{Green:1987sp}.
The supersymmetric {gluon} vertex operator   in the
light-cone gauge 
{reads}
\begin{equation}
    V^{\rm vector}(\epsilon,k)=\int dt :i \epsilon^j
      \left(X'^j(t)-{\sqrt{\alpha'}k_l \over4} \gamma^{jl}_{ab}S^a(t)S^b(t) 
      \right)\, e^{ik\cdot X(t)}:.
    \end{equation}
    
At leading order in $\alpha'$, the commutator between two
supersymmetric vertex vertex operators 
{gives}
\begin{multline}
  [V^{\rm vector}(\epsilon_1,k_1),V^{\rm
  vector}(\epsilon_2,k_2)=f^{[1]}_i   \tilde E^{{\rm
    vector},i}(k_1+k_2)+f^{[2]}_{ij} \tilde E^{{\rm
    tensor},ij}(k_1+k_2)\cr+
{\alpha'\over4} F^1_{ij} F^2_{lk} \tilde E^{{\rm tensor},ij,kl}(k_1+k_2),
\end{multline}
where  the structure constants  $f^{[1]}_i$ and $f^{[2]}_{ij}$  are the
same as in~\eqref{e:fcste} evaluated 
{in} 
the transverse directions
$1\leq i,j\leq D-1$, 
{with} 
the generators
{given by}
\begin{multline}
   \tilde E^{{\rm
    vector},i}(k_1+k_2)= \int_{0}^{1}dt_1 \int_{\mathcal C } dt_{2}
 (t_1-t_{2})^{\alpha' k_{1}\cdot
   k_{2}-1}\cr\times:i  \left(X'^j(t_2)-{\sqrt{\alpha'}k_l \over4} S(t_2)\gamma^{jl}S(t_2) 
      \right) e^{ik_{1}\cdot
   X(t_{1})}e^{ik_{2}\cdot
   X(t_{2})}:,
\end{multline}
and
\begin{multline}
\tilde E^{\text{tensor},ij}(k_{1}+k_{2}):=
\int_{0}^{1}dt_1 \int_{\mathcal C } dt_{2}
(t_1-t_{2})^{\alpha' k_{1}\cdot k_{2}}\cr
:i  \left(X'^i(t_2)-{\sqrt{\alpha'}k_l \over4} S(t_2)\gamma^{il}S(t_2) 
    \right) \,i  \left(X'^j(t_2)-{\sqrt{\alpha'}k_r \over4} S(t_2)\gamma^{jr}S(t_2) 
     \right)e^{ik_{1}\cdot
   X(t_{1})}e^{ik_{2}\cdot
   X(t_{2})}:,
\end{multline}
which are the supersymmetric generalisation of the generators
in~\eqref{eq:4pt-vector} and~\eqref{eq:4pt-tensor}.
And 
{we have a}
new tensor generator
\begin{equation}
    \tilde E^{{\rm tensor},ij,kl}(k_1+k_2):=\int_{0}^{1}dt_1 \int_{\mathcal C } dt_{2}
 (t_1-t_{2})^{\alpha' k_{1}\cdot
   k_{2}-1}\,  :S\gamma^{ij}\,\gamma^{kl}S: \,  e^{ik_{1}\cdot
   X(t_{1})}e^{ik_{2}\cdot
   X(t_{2})}:.
\end{equation}

\medskip

The numerator factor for the emission of gluons from a massive fermion line, is given by  the expectation value of the nested  commutator of gluon vertex operators, this time evaluated between the fermionic external states,
\begin{multline}
 \mathcal N^{\rm fermions}(1,\sigma(2),\dots,\sigma(n)) 
=\frac{1}{n} \left(\frac{-i}{ \pi } \right)^{n-1}
\lim_{\alpha'\to0}
\alpha'^{-2n+1}\cr
\Bigl\langle u(p)\Bigr|\left[\left[\left[V^{\textrm{vector}}_{1},V^{\textrm{vector}}_{\sigma(2)}\right]_{\alpha'},V^{\textrm{vector}}_{\sigma(3)}\right]_{\alpha'},\dots,V^{\textrm{vector}}_{\sigma(n)}\right]_{\alpha'}\Bigl|u(-p') \Bigr\rangle .
\end{multline}
{where the}
fermions
{are defined by}  
vertex operators 
{of the form}
~\cite{Green:1987sp} 
\begin{equation}
    V^{\rm fermion}(p)=\int dt :\left(\sqrt{p^+\over2} u^a
      S^a(t)+{u^{\dot a}\over \sqrt{2p^+}}\left((\gamma\cdot X'
          S)^{\dot a}+{\sqrt{\alpha'}\over12} (\gamma^i S)^{\dot a}  \gamma^{ij}_{cd}S^c S^d p_j\right)\right)\, e^{i p\cdot X}:.
\end{equation}
We consider a massive fermion, which can be obtained by thinking that
some of the ten dimensions are compactified so that $p^2\neq0$.

The numerator factor  for the emission of two gluons from a line of
massive fermions is given by
\begin{equation}
\mathcal N^{\rm fermions}(1,2) 
=\frac{1}{2} \left(\frac{-i}{ \pi } \right)
\lim_{\alpha'\to0}
\alpha'^{-3} \Bigl\langle u(p)\Bigr|\left[V^{\textrm{vector}}_{1},V^{\textrm{vector}}_{2}\right]_{\alpha'}\Bigl|u(-p') \Bigr\rangle .
\end{equation}
The evaluation of the expectation value follows the derivation 
of the two fermions, two gluons open string amplitude computed in
section~7.4.2 of~\cite{Green:1987sp}. 
Using the gauge~\eqref{e:barepsilon} for the gluon 1, 
we  obtain for the numerator factor in a covariant form
\begin{equation}
\mathcal N^{\rm fermion}(1,2)={1\over k_1\cdot p}\bar u(p)\left( \gamma_\mu \, p_\nu (F_1^{\nu})_\rho F_2^{\rho\mu}+ \frac14 [\gamma_\mu,\gamma_\nu] F_1^{\mu\nu} \gamma^\rho \, p_1^\lambda (F_2)_{\lambda\rho}\right)u(-p') \, .
\end{equation}
This expression  matches equation~(2.9) of~\cite{Bjerrum-Bohr:2024fbt}. 

The extension to the multi-gluon emission is immediate as this follows
from the kinematic algebra detailed above adapted to the
supersymmetric case.


\section{Calculating HEFT numerators}
\label{sec:numerator-derivation}

In this section, we describe  an algorithmic way of computing the HEFT
numerator factors for the emission of gluons from a massive scalar line. A \texttt{Mathematica} code  for generating the
numerator factor and a worksheet illustrating the algorithm are
available  at
this github repository~\href{https://github.com/Yi-hongWang/Stringy-Numerator}{https://github.com/Yi-hongWang/Stringy-Numerator}.

\subsection{The algorithmic implementation}
\label{sec:algorithm}

This section details the algorithm used to calculate the 
numerator factors.  We calculate iteratively the nested commutator $[[E_{k_{1}}^{\text{vector}},E_{k_{2}}^{\text{vector}}]_{\alpha'},\dots,E_{k_{n}}^{\text{vector}}]_{\alpha'}$,
adding one gluon at a time. Generically we need to consider the OPE
of an $\ell$-th gluon $E_{k_{\ell}}^{\text{vector}}\left(\epsilon_{\ell},k_{\ell}\right)=\int dz_{\ell}:i\epsilon_{\ell}\cdot X^{(1)}(z_{\ell})e^{ik_{\ell}\cdot X(z_{\ell})}:$
with the nested commutator
\begin{equation}
  [[E_{k_{1}}^{\text{vector}},E_{k_{2}}^{\text{vector}}]_{\alpha'},\dots,E_{k_{\ell-1}}^{\text{vector}}]_{\alpha'},
  \end{equation}
which we assume to take the following general form similar to~(\ref{eq:4pt-vector})
and~(\ref{eq:4pt-tensor}).

\begin{equation}\label{eq:lm1}
  \int_{T^{\ell-2}}\prod_{i=1}^{\ell-1}dt_{i}\sum_{\alpha}g_{\alpha}\left(k_{1},k_{2},F_{2}\dots k_{n-1},F_{n-1}\right)
  \times I_{\alpha}\left(t_{1}\dots t_{n-1}\right)O_{\alpha}\text{\ensuremath{\prod_{1\leq j<i\leq\ell-1}\left(t_{i}-t_{j}\right)^{\alpha' k_{i}\cdot k_{j}}}}.
\end{equation}
{In the above equation, we} 
collect momentum-dependent coefficients into $g_{\alpha}$
and world-sheet variable-dependent coefficients
into $I_{\alpha}$ and sum over the various contributions  labelled by $\alpha$.
The operator $O_{\alpha}$ is generically a higher-rank tensor given by the following product 
\begin{equation}
:\prod_{j\in R_{\alpha}}\mathcal T_{j}\cdot X'\left(t_{j}\right)\prod_{i=1}^{\ell-1}e^{ik_{i}\cdot X\left(t_{i}\right)}:,\label{eq:tlm1}
\end{equation}
over the set $R_\alpha$ of indices and 
the tensor  $\mathcal T_{j}$ is of the form $\epsilon_{*}\cdot F_{*}\cdots F_{j}$ as discussed in section~\ref{sec:queues}

\medskip
The \texttt{Mathematica} code  enumerates all possible contractions of~(\ref{eq:tlm1}) with the
$\ell$-th gluon. We have the following three different
scenarios when a subset $Q\subset R_{\alpha}$ of the tensor indices
contracts with the new vertex operator $:~\epsilon_{\ell}\cdot X'\left(t_{\ell}\right)e^{ik_{\ell}\cdot X\left(t_{\ell}\right)}~:$
\begin{enumerate}
\item[(a)]  One of the vectors $\mathcal T_{a}\cdot X'\left(t_{a}\right)$, $a\in Q$,
contracts with $\epsilon_{\ell}\cdot X'\left(t_{\ell}\right)$ while
the rest contract with $e^{ik_{\ell}\cdot X\left(t_{\ell}\right)}$.
\item[(b)]  All vectors in $Q$ contract with $e^{ik_{\ell}\cdot X\left(t_{\ell}\right)}$
while $\epsilon_{\ell}\cdot X'\left(t_{\ell}\right)$ contract with
$e^{ik_{b}\cdot X\left(t_{b}\right)}$, $b\in\{1,\dots,\ell\}\backslash Q$.
\item[(c)] All vectors in $Q$ contract with $e^{ik_{\ell}\cdot X\left(t_{\ell}\right)}$
while $\epsilon_{\ell}\cdot X'\left(t_{\ell}\right)$ left uncontracted. 
\end{enumerate}
Let $q$ be the  element with the greatest index in $Q$, so the last term in (a)
reads
\begin{equation}
 \left(-1\right)^{\left|Q\right|-1}\frac{\epsilon_{\ell}\cdot \mathcal T_{q}}{\left(t_{\ell}-t_{a}\right)^{2}}\prod_{i\in Q\backslash\{q\}}\frac{\mathcal T_{i}\cdot k_{\ell}}{t_{\ell}-t_{i}}
  \times:\prod_{j\in R\backslash Q}\mathcal T_{j}\cdot X'\left(t_{j}\right)\prod_{i=1}^{n+1}e^{ik_{i}\cdot X\left(t_{i}\right)}:\prod_{1\leq j<i\leq\ell}\left(t_{i}-t_{j}\right)^{\alpha' k_{i}\cdot k_{j}}.
\end{equation}
 Converting all the polarisation dependence to field-strength as in example~\eqref{e:V1vecV2vec} requires making some choice in applying the  IBP, which complicates the algorithmic procedure. Therefore, we  only perform IBP on the double pole of this term with respect to the new variable $t_{\ell}$. 

The derivative may then act on the simple
poles of $t_{\ell}$, the operators, and the Koba-Nielsen factor. Each type
could be combined with terms in (a), (b), (c) respectively and rewritten
in terms of $F_{\ell}$:
\begin{enumerate}
\item[(a')]  Differentiation on the simple poles combines with other terms in
(a) into the summation
\begin{multline}
\left(-1\right)^{\left|Q\right|}  \sum_{i\in Q\backslash\{a\}}\frac{\mathcal T_{a}\cdot F_{\ell}\cdot \mathcal T_{q}}{\left(t_{\ell}-t_{a}\right)^{2}}\prod_{i\in Q\backslash\{a,q\}}\frac{\mathcal T_{i}\cdot k_{\ell}}{t_{\ell}-t_{i}}
 \times :\prod_{j\in R\backslash Q}\mathcal T_{j}\cdot X'\left(t_{j}\right)\prod_{i=1}^{n+1}e^{ik_{i}\cdot X\left(t_{i}\right)}:\cr\times\prod_{1\leq j<i\leq\ell}\left(t_{i}-t_{j}\right)^{\alpha' k_{i}\cdot k_{j}}.
\end{multline}
\item[(b')]  Differentiation on the normal ordered operators leads to a new tensor
term which combines with (b) into
\begin{multline}
\left(-1\right)^{\left|Q\right|}  \frac{1}{\left(t_{\ell}-t_{q}\right)}\prod_{i\in Q\backslash\{q\}}\frac{\mathcal T_{i}\cdot k_{\ell}}{t_{\ell}-t_{i}}:\mathcal T_{q}\cdot F_{\ell}\cdot X'\left(t_{\ell}\right):
 \times :\prod_{j\in R\backslash Q}\mathcal T_{j}\cdot X'\left(t_{j}\right)\prod_{i=1}^{n+1}e^{ik_{i}\cdot X\left(t_{i}\right)}:\cr\times\prod_{1\leq j<i\leq\ell}\left(t_{i}-t_{j}\right)^{\alpha' k_{i}\cdot k_{j}}.
\end{multline}
\item[(c')]  Differentiation on Koba-Nielsen factors combines with (c) into 
\begin{multline}
\left(-1\right)^{\left|Q\right|-1}  \left(\sum_{b\in\left[n\right]\backslash Q}\frac{k_{b}\cdot F_{l}\cdot \mathcal T_{q}}{\left(t_{\ell}-t_{b}\right)\left(t_{\ell}-t_{q}\right)}\right)\prod_{i\in Q\backslash\{q\}}\frac{\mathcal T_{i}\cdot k_{\ell}}{t_{\ell}-t_{i}}
  \cr\times :\prod_{j\in R\backslash Q}\mathcal T_{j}\cdot X'\left(t_{j}\right)\prod_{i=1}^{n+1}e^{ik_{i}\cdot X\left(t_{i}\right)}:\prod_{1\leq j<i\leq\ell}\left(t_{i}-t_{j}\right)^{\alpha' k_{i}\cdot k_{j}}.
\end{multline}
\end{enumerate}
The main function in our code is programmed to enumerate  all terms of the three types listed above for any tensor and any subset $Q$. We then enumerate and sum over all possible subsets for all terms carried over from previous calculations. When the terms are collected, the result will naturally be in the same form as in~(\ref{eq:lm1}) and available for the next step of iteration. To obtain the $n$ gluon numerator, we carry on with this iteration until the $n$-th fold, after which we calculate its OPE with the scalar vertex operator $:e^{ip\cdot X\left(t_{n+1}\right)}:$, keeping only the terms that satisfy the ascending order criteria
{(explained in section \ref{sec:integrals}),} 
and then evaluate them by the integration rule~(\ref{eq:integral-rules}).

\subsubsection{Example: contractions  for the four gluons emission}

We illustrate the algorithm with the derivation of the four gluons emission case.


We add the fourth gluon {vertex operator} $:\epsilon_{4}\cdot X'(t_{4})e^{ik_{4}\cdot X(t_{4})}:$
to the nested commutators. Let us focus on just the tensor term
\begin{equation}:\frac{\alpha'\bar\epsilon_{1}\cdot F_{3}\cdot X'\left(t_{3}\right)\, \epsilon_{2}\cdot X'\left(t_{2}\right)}{\left(t_{3}-t_{1}\right)}\prod_{1\leq i<j\leq3}\left(t_{i}-t_{j}\right)^{\alpha' k_{i}\cdot k_{j}}  :\prod_{i=1,2,3}e^{k_{i}\cdot X'\left(t_{i}\right)}:,
\end{equation} 
carried over
{from previous iterations.} 
In the case where only $\epsilon_{2}\cdot X'\left(t_{2}\right)$ is contracted, we have the following four  terms contributing in leading order in $\alpha'$
\begin{multline}
  \underset{(0)}{\underbrace{\alpha'^{2}\frac{\epsilon_{2}\cdot\epsilon_{4}}{\left(t_{4}-t_{2}\right)^{2}}\frac{\bar\epsilon_{1}\cdot F_{3}\cdot X'\left(t_{3}\right)}{\left(t_{3}-t_{1}\right)}}},\quad\underset{(1)}{\underbrace{\alpha'^{3}\frac{\epsilon_{4}\cdot k_{1}}{\left(t_{4}-t_{1}\right)}\frac{\epsilon_{2}\cdot k_{4}}{\left(t_{4}-t_{2}\right)}\frac{\bar\epsilon_{1}\cdot F_{3}\cdot X'\left(t_{3}\right)}{\left(t_{3}-t_{1}\right)}}},\cr
  \underset{(2)}{\underbrace{\alpha'^{3}\frac{\epsilon_{4}\cdot k_{3}}{\left(t_{4}-t_{3}\right)}\frac{\epsilon_{2}\cdot k_{4}}{\left(t_{4}-t_{2}\right)}\frac{\bar\epsilon_{1}\cdot F_{3}\cdot X'\left(t_{3}\right)}{\left(t_{3}-t_{1}\right)}}},\quad\underset{(3)}{\underbrace{\alpha'^{2}\frac{\epsilon_{2}\cdot k_{4}}{\left(t_{4}-z_{2}\right)}\frac{\bar\epsilon_{1}\cdot F_{3}\cdot X'\left(t_{3}\right)}{\left(t_{3}-t_{1}\right)}\epsilon_{4}\cdot X'\left(t_{4}\right)}}
\end{multline}
Performing the IBP on the term $(0)$ and combining it with the other three terms,  we get
\begin{multline}
  \underset{(1')}{\underbrace{\alpha'^{3}\frac{\epsilon_{2}\cdot F_{4}\cdot k_{1}}{\left(t_{4}-t_{2}\right)\left(t_{4}-t_{1}\right)}\frac{\bar\epsilon_{1}\cdot F_{3}\cdot X'\left(t_{3}\right)}{\left(t_{3}-t_{1}\right)}}},\quad\underset{(2')}{\underbrace{\alpha'^{3}\frac{\epsilon_{2}\cdot F_{4}\cdot k_{3}}{\left(t_{4}-t_{2}\right)\left(t_{4}-t_{3}\right)}\frac{\bar\epsilon_{1}\cdot F_{3}\cdot X'\left(t_{3}\right)}{\left(t_{3}-t_{1}\right)}}},\cr
  \underset{(3')}{\underbrace{\alpha'^{2}\frac{\bar\epsilon_{1}\cdot F_{3}\cdot X'\left(t_{3}\right)}{\left(t_{3}-t_{1}\right)}\frac{\epsilon_{2}\cdot F_{4}\cdot X'\left(t_{4}\right)}{\left(t_{4}-t_{2}\right)}}}.
\end{multline}
Its contribution to the numerator factor is given by contracting with $e^{ip\cdot X\left(t_{5}\right)}$ and imposing the gauge choice $\bar\epsilon_{1}\cdot p=0$ along with the ascending order criteria 
{explained in section \ref{sec:integrals}.} 
Only {the} term $(3')$ 
{then}
survives and leads to the integrand 
\begin{equation}
\alpha'^{4}\frac{\bar{\epsilon}_{1}\cdot F_{3}\cdot p\,\epsilon_{2}\cdot F_{4}\cdot p}{\left(t_{3}-t_{1}\right)\left(t_{5}-t_{3}\right)\left(t_{4}-t_{2}\right)\left(t_{5}-t_{4}\right)}.
\end{equation}

In the case where both  $\bar\epsilon_{1}\cdot F_{3}\cdot\dot{X}\left(t_{3}\right)$
and $\epsilon_{2}\cdot\dot{X}\left(t_{2}\right)$ are contracted, the
leading {order}
terms in $\alpha'$ are 
\begin{align}
 & \underset{(4)}{\underbrace{\alpha'^{3}\frac{\bar\epsilon_{1}\cdot F_{3}\cdot k_{4}\epsilon_{2}\cdot k_{4}}{\left(t_{3}-t_{1}\right)\left(t_{4}-t_{2}\right)\left(t_{4}-t_{3}\right)}\epsilon_{4}\cdot X'\left(t_{4}\right)}},\quad\underset{(5)}{\underbrace{\alpha'^{3}\frac{\bar\epsilon_{1}\cdot F_{3}\cdot k_{4}\epsilon_{2}\cdot\epsilon_{4}}{\left(t_{3}-t_{1}\right)\left(t_{4}-t_{2}\right)^{2}\left(t_{4}-t_{3}\right)}}},\cr
 & \underset{(6)}{\underbrace{\alpha'^{4}\frac{\epsilon_{2}\cdot k_{4}\epsilon_{4}\cdot k_{1}\bar\epsilon_{1}\cdot F_{3}\cdot k_{4}}{\left(t_{3}-t_{1}\right)\left(t_{4}-t_{1}\right)\left(t_{4}-t_{2}\right)\left(t_{4}-t_{3}\right)}}},\quad\underset{(7)}{\underbrace{\alpha'^{3}\frac{\bar\epsilon_{1}\cdot F_{3}\cdot\epsilon_{4}\epsilon_{2}\cdot k_{4}}{\left(t_{3}-t_{1}\right)\left(t_{4}-t_{2}\right)\left(t_{4}-t_{3}\right)^{2}}}}.
\end{align}
Performing the IBP on {the} term $(7)$ leads to
 {contributions that when added with the} terms
 $(4)$, $(5)$ and $(6)$ 
{lead to an expression} in terms of $F_{i}$'s:
\begin{align}
\underset{(4')}{\underbrace{\alpha'^{3}\frac{\epsilon_{2}\cdot k_{4}}{\left(t_{4}-t_{2}\right)}\frac{\bar\epsilon_{1}\cdot F_{3}\cdot F_{4}\cdot X'\left(t_{4}\right)}{\left(t_{3}-t_{1}\right)\left(t_{4}-t_{3}\right)}}},\quad\underset{(5')}{\underbrace{-\alpha'^{3}\frac{\bar\epsilon_{1}\cdot F_{3}\cdot F_{4}\cdot\epsilon_{2}}{\left(t_{3}-t_{1}\right)\left(t_{4}-t_{2}\right)^{2}\left(t_{4}-t_{3}\right)}}},\cr
\underset{(6')}{\underbrace{\alpha'^{4}\frac{\epsilon_{2}\cdot k_{4} \,\bar\epsilon_{1}\cdot F_{3}\cdot F_{4}\cdot k_{1}}{\left(t_{3}-t_{1}\right)\left(t_{4}-t_{1}\right)\left(t_{4}-t_{2}\right)\left(t_{4}-t_{3}\right)}}}.
\end{align}
However, none of the above terms passes the ascending order criteria
after contraction with the last scalar vertex operator, and therefore 
 {do not contribute} to  {the} leading {order} contribution in the limit $\alpha'\rightarrow0$. 

\medskip

{For simplicity we may set the reference momentum for gluon $2$ to be $p$, so that $p\cdot \epsilon_2=0$.}
The remaining two cases are those where the contracting set is $\{\bar\epsilon_{1}\cdot F_{3}\cdot X'\left(t_{3}\right)\}$
or the empty set, and can be calculated similarly. Contracting with  $\{\bar\epsilon_{1}\cdot F_{3}\cdot X'\left(t_{3}\right)\}$
leads to no leading contribution in $\alpha'$; they do not contribute to the field theory  final result.  The empty contracting set leads to the following two terms with $i=1,2,3$
\begin{equation}
    \alpha' \frac{:\bar\epsilon_{1}\cdot F_{3}\cdot X'\left(t_{3}\right)\epsilon_{2}\cdot X'\left(t_{2}\right)\epsilon_{4}\cdot X'\left(t_{4}\right):}{\left(t_{3}-t_{1}\right)},\quad 
    \alpha'^{2} \frac{k_{i}\cdot\epsilon_{4}}{\left(t_{4}-t_{i}\right)}\frac{:\bar\epsilon_{1}\cdot F_{3}\cdot X'\left(t_{3}\right)\epsilon_{2}\cdot X'\left(t_{2}\right):}{\left(t_{3}-t_{1}\right)}.
\end{equation} 
The only possible contractions of this operator are with the massive (tachyonic) vertex operator, which leads to
\begin{equation}
    \alpha'^{4}\frac{\bar\epsilon_{1}\cdot F_{3}\cdot p\,\epsilon_{2}\cdot p\,\epsilon_{4}\cdot p}{\left(t_{3}-t_{1}\right)(t_3-1)(t_2-1)(t_4-1)},\quad
    \alpha'^{4} \frac{k_{i}\cdot\epsilon_{4}}{\left(t_{4}-t_{i}\right)}\frac{\bar\epsilon_{1}\cdot F_{3}\cdot p\,\epsilon_{2}\cdot p}{\left(t_{3}-t_{1}\right)(t_3-1)(t_2-1)}.
\end{equation}

\subsection{Integrations over the position of the vertex operators}
\label{sec:integrals}

\begin{figure}[h]
\centering
\includegraphics[width=3cm]{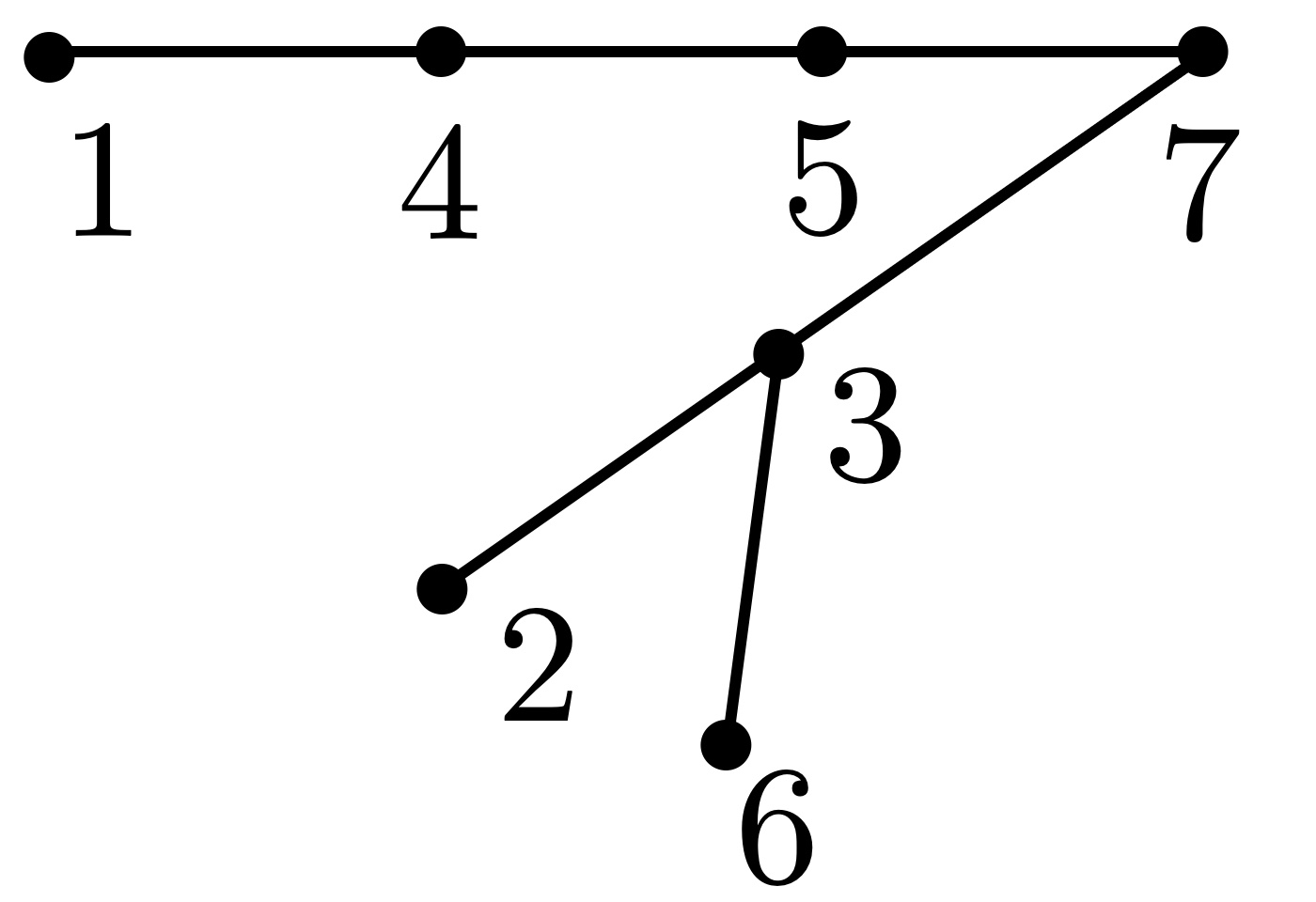}
\hspace{0.5cm} 
\raisebox{1cm}{$ \sim $} 
\hspace{0.5cm}
\includegraphics[width=3cm]{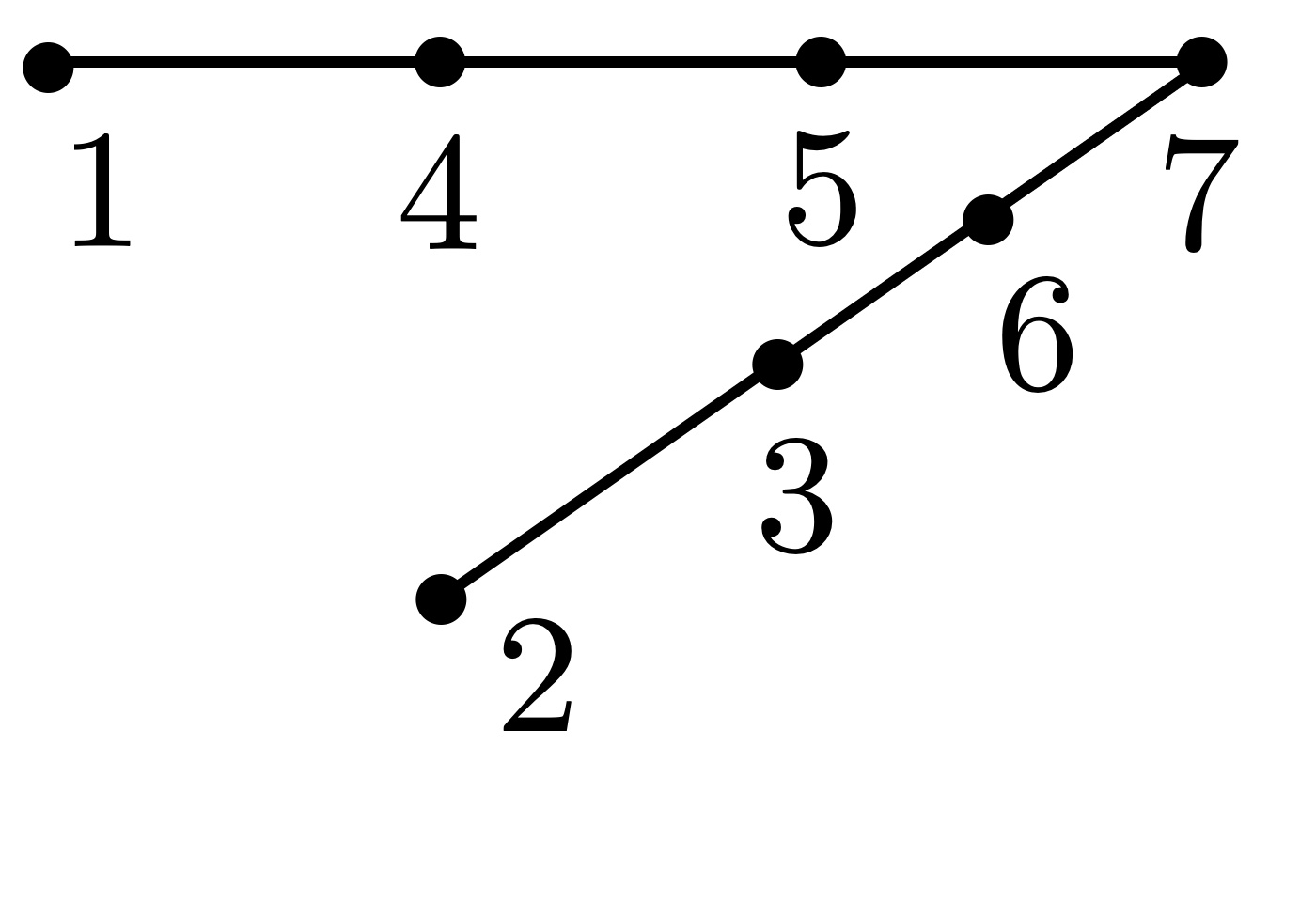}
\caption{Removing subtrees using partial fraction}
\label{fig:IBPgraph}
\end{figure}

We explain how to integrate over the positions of the vertex operators.

For the moment, we concentrate on the integrals after taking the
expectation value, so that  no operators are left in the integrand. By introducing a graphical notation of~\cite{Baadsgaard:2015voa,He:2018pol}, where we draw a line connecting the points $i$ and $j$ to represent a pole $1/(t_{i}-t_{j})$ multiplied by the Koba-Nielsen factor $(t_i-t_j)^{\alpha' k_i\cdot k_j}$ in the integrand, we are interested in the vanishing $\alpha'$ limit of their integrals over enclosing unit circles. In practice, this can be obtained using their original definition as 
{product of }
the momentum kernel  matrix with 
{ordered}
line interval integrals, which in turn can be computed using the integration rules of~\cite{Baadsgaard:2015voa,Lam:2015sqb}.

\begin{figure}[h]
    \centering
    \includegraphics[width=3cm]{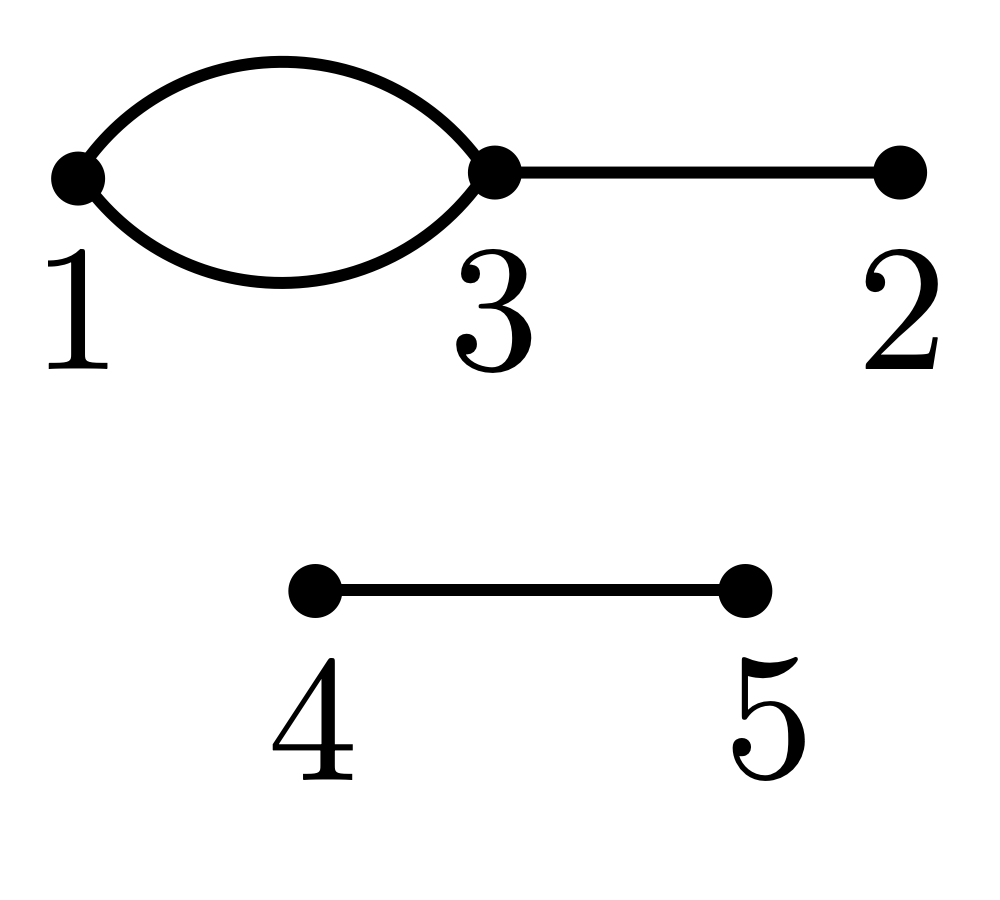}\hspace{0.5cm}
     \includegraphics[width=3cm]{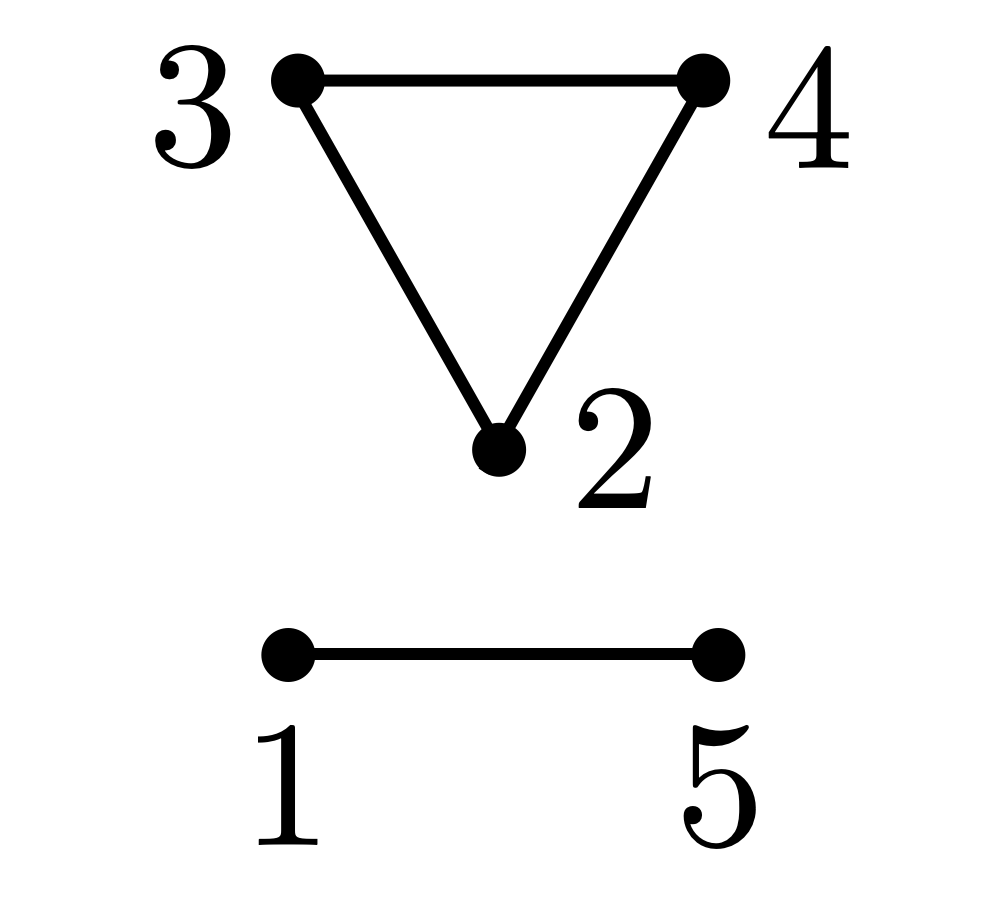}\hspace{0.5cm}
      \includegraphics[width=3cm]{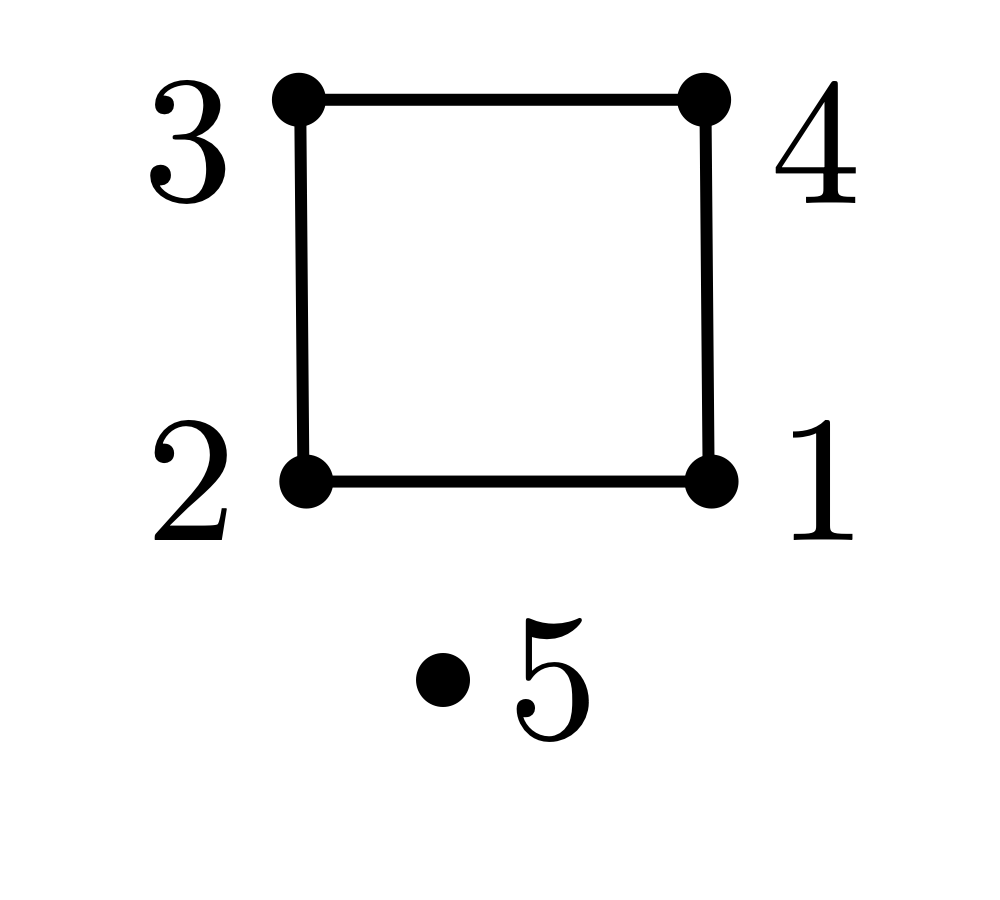}
    \caption{Disconnected graphs with loops}
    \label{fig:loop}
\end{figure}

First, note that graphs with loops are related to tree graphs by IBP~\cite{Schlotterer:2016cxa,He:2018pol}, which brings  powers of $\alpha'$ from the Koba-Nielsen factor. Therefore, graphs with loops are subleading. By counting the number of poles, we deduce that unconnected graphs  always have loops, and are therefore suppressed by powers of $\alpha'$. Finally, we remark  that subtrees can be transformed into sums of a single line by repeated use of partial fractions, with the largest and smallest number on that line at either end, as represented in figure~\ref{fig:IBPgraph}.

\begin{figure}[h!]
        \centering
        \includegraphics[width=5cm]{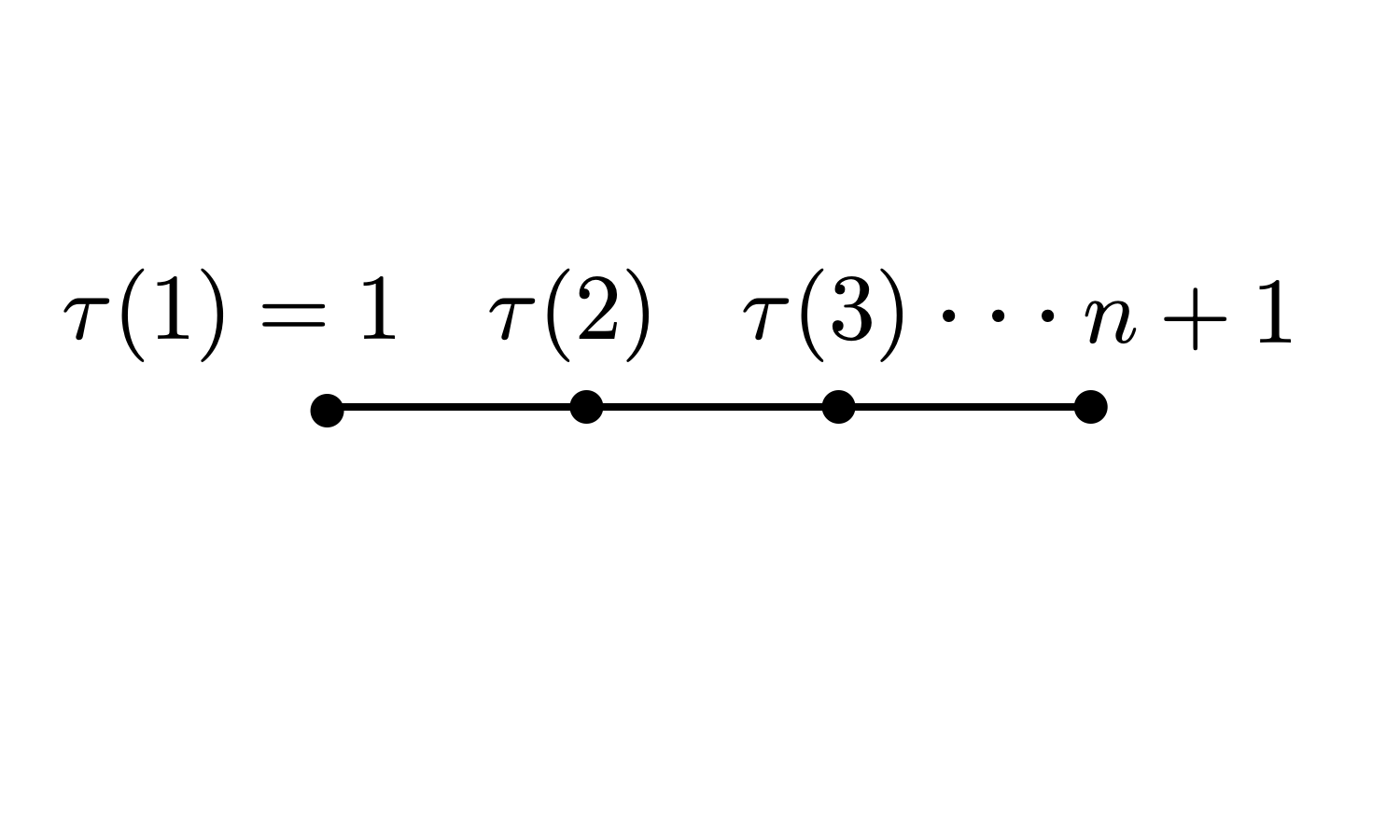}\hspace{2cm}
     \includegraphics[width=5cm]{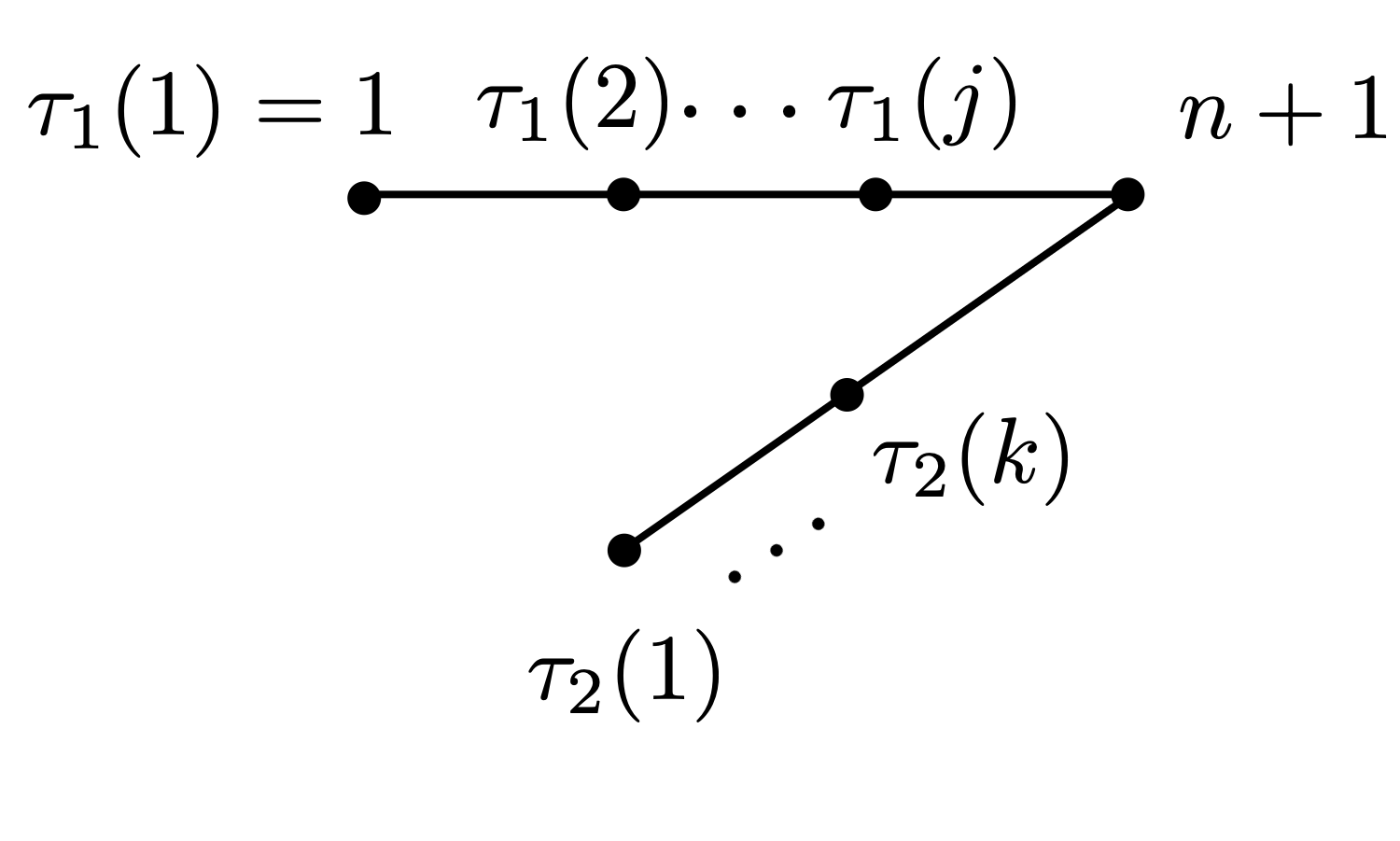}
         \caption{Graphs representing the pole structure of integrands}
         \label{fig:ascending-order}
\end{figure}

Taking all these into account, and considering the position of the
massive particle labelled  $n+1$ as the root, we only need to consider
multiple  straight lines rooted at  $n+1$, as represented in
figure~\ref{fig:ascending-order}.

A direct application of the integration rules shows that the product of the momentum kernel matrix with a single straight line with the order of the labels  $\{\tau(1)=1, \tau(2), \tau(3), \dots,\tau(n),n+1\}$ gives $1$ if $\tau$ is an ascending order, i.e. $\tau(i+1)>\tau(i)$, and vanishes otherwise. 
Similarly, if there are multiple lines rooted at $n+1$,  one multiplies the result of each line. Therefore, the result vanishes unless each line is in relatively ascending order.

For a graph with $m$ lines rooted at $n+1$, the contribution is given by the following product of $(m-1)$ factors,
one for each line except the line that containing the label  $1$.
For each line, we associate the factor~\cite{Brandhuber:2021bsf}
\begin{equation}
\frac{k_{\theta(\tau_i(1))}\cdot k_{\tau_i(1)}}{ k_{\Sigma(\tau_i(1))}\cdot p},
\end{equation}
where $k_{\theta(\tau_i(1))}$ in the numerator is the sum of all the momenta $k_r$ that  appear to the left of $\tau_{i}(1)$  on the line 
with the extra condition that $r$ has to be less than $\tau_{i}(1)$,
and in the denominator $k_{\Sigma(\tau_i(1))}$  is the sum of all the momenta $k_r$ that  appear to the left of $\tau_{i}(1)$  on the line without restrictions.

We then sum over the permutations of the result of $(m-1)$ lines giving
\begin{equation}
\sum_{\rho\in \mathfrak S_{m-1}}\prod_{i=2}^{m}\frac{k_{\theta(\tau_{\rho(i)}(1))}\cdot k_{\tau_{\rho(i)}(1)}}{k_{\text{all left to }\tau_{\rho(i)}(1)}\cdot p}.\label{eq:integral-rules}
\end{equation}
Note that power counting shows that a $p$  in the denominator
leads to a $\alpha'^{\frac{1}{2}}$ increase in order, so that for
the single line graph we have $\mathcal{O}(\alpha'^{0})$ (numerically,
it is $1$ or $0$), while a $m$-tuple line graph contributes $\mathcal{O}(\alpha'^{\frac{m-1}{2}})$.

\medskip
Here we will illustrate the derivation with the following seven-gluon example. The tachyonic massive particle vertex operators are at position 
$t_8=1$ and $t_9=\infty$, and the gluon 1 at position $t_1=0$, so that the integrand of the nested commutator reads
\begin{equation}
{1\over (t_1-t_4)(t_4-t_6)(t_6-t_8)}
\times {1\over (t_2-t_8)}
\times {1\over (t_3-t_5)(t_5-t_7)(t_7-t_8)}\prod_{1\leq i<j\leq 8}\left(t_{j}-t_{i}\right)^{\alpha'k_{i}\cdot k_{j}}.\label{eq:integral-example}
\end{equation}
We needed to compute the 
leading order {contribution in $\alpha'$} of the product of $(n-1)$ basis momentum kernel matrix
with ordered domain world-sheet integrals. The leading contribution
from the world-sheet integral alone can be readily read off using
the integration rules from~\cite{Baadsgaard:2015voa,Lam:2015sqb}, where it is shown to arise from collisions of adjacent world-sheet variables.

\subsubsection{Example: Integrations rules for the seven gluons case}
We give a demonstration of the integration rules for the seven gluons
case.

For example the integrand~(\ref{eq:integral-example})  gives  non-vanishing results only on the domains
\begin{equation}
0=t_{1}<t_{4},t_{6}<t_{3},t_{5},t_{7}<t_{2}<t_{8}=1.\label{eq:domain2}
\end{equation}
and
\begin{equation}
0=t_{1}<t_{4},t_{6}<t_{2}<t_{3},t_{5},t_{7}<t_{8}=1.\label{eq:domain1}
\end{equation}
The rest of the ordered domain simply does not have enough adjacent
world-sheet variables that match the number of poles  {appearing} in the integrand. 

For the ordered domain of the integration $0~=t_{1}<t_{4}<t_{6}<t_{3}<t_{7}<t_{5}<t_{2}<t_{8}=1$
inside~(\ref{eq:domain2}), the $\alpha'$ leading order of the integration
is given by
\begin{equation}
\alpha'^{-6}\,\frac{m\left(146\left|146\right.\right)m\left(357\left|375\right.\right)m\left(2\left|2\right.\right)}{\left(k_{1}+k_{4}+k_{6}\right)\cdot p\left(k_{1}+k_{3}+k_{4}+k_{5}+k_{6}+k_{7}\right)\cdot p},
\end{equation}
where 
\begin{align}
m\left(146\left|146\right.\right)&=
\frac{1}{k_{1}\cdot k_{4}+k_{1} \cdot k_{6}+k_{4}\cdot k_{6} } 
\left( \frac{1}{k_{1}\cdot k_{4}} +
\frac{1}{k_{4}\cdot k_{6}} \right),\cr
m\left(357\left|375\right.\right)&=
\frac{1}{k_{3}\cdot k_{5}+k_{3}\cdot k_{7}+k_{5}\cdot k_{7} } 
\left( \frac{-1}{k_{5}\cdot k_{7}}
\right),
\cr
m\left(2\left|2\right.\right)=1.
\end{align}
are the off-shell field theory propagator matrices~\cite{Cachazo:2013iea,Mafra:2016ltu}.
{Similarly}  
for the integrals for all other ordered domains.

We are interested in the product of the momentum kernel with ordered integrals.
In this matrix product, the integral~(\ref{eq:integral-example})
is multiplied by the following entry of the string theory moment kernel
matrix $\mathcal{S}_{\alpha'}[1234567|1463752]$, 
its $\alpha'$ leading contribution is simply the field theory moment
kernel $\mathcal{S}[1234567|1463752]$. Carrying out the same calculation
for all other terms in the matrix product corresponding to the domains
in~(\ref{eq:domain2}), and we arrive at the following sum
\begin{equation}
\frac{\sum_{\sigma_{1}\in \mathfrak S_{2}}\sum_{\sigma_{2}\in \mathfrak S_{3}}\mathcal{S}[1234567|1\sigma_{1}\left(46\right)\sigma_{2}\left(375\right)2]|_p\,m\left(146\left|1\sigma_{1}\left(46\right)\right.\right)m\left(357\left|\sigma_{2}\left(357\right)\right.\right)m\left(2\left|2\right.\right)}{\alpha'^{6}\left(k_{1}+k_{4}+k_{6}\right)\cdot p\left(k_{1}+k_{3}+k_{4}+k_{5}+k_{6}+k_{7}\right)\cdot p},\label{eq:domain2sum}
\end{equation}
where $\sigma_{1}$ is a permutation of the indices $\left\{ 4,6\right\} $,
and $\sigma_{2}$ is a permutation of the indices $\left\{ 3,5,7\right\} $
respectively. The above expression  can be further simplified by repeatedly  using  the KK-relation~\cite{Kleiss:1988ne} 
\begin{eqnarray}
&& m\left(12\dots\ell\left|\ell12\dots\ell-1\right.\right)+m\left(12\dots\ell\left|1\ell2\dots\ell-1\right.\right) 
+m\left(12\dots\ell\left|12\ell\dots\ell-1\right.\right) 
\label{eq:KK}\cr
&&
+ \dots+m\left(12\dots\ell\left|12\dots\ell-1\ell\right.\right)=0,
\end{eqnarray}
 and the off-shell BCJ-relations~\cite{Du:2011js}
\begin{eqnarray}
&&
\left(k_{1}\cdot k_{\ell}\right)m\left(12\dots\ell\left|1\ell2\dots\ell-1\right.\right)+\left(k_{1}\cdot k_{\ell}+k_{2}\cdot k_{\ell}\right)m\left(12\dots\ell\left|12\ell\dots\ell-1\right.\right) 
\cr
&& +\dots +\left(k_{1}\cdot k_{\ell}+k_{2}\cdot k_{\ell}\dots+k_{\ell-1}\cdot k_{\ell}\right)m\left(12\dots\ell\left|12\dots\ell-1\ell\right.\right)   \cr
&& =m\left(12\dots\ell-1\left|12\dots\ell-1\right.\right), \label{eq:BCJ}
\end{eqnarray}
and
\begin{eqnarray}
&&  \left(k_{1}\cdot k_{i}\right)m\left(12\dots\ell\left|1i2\dots\ell\right.\right)+\left(k_{1}\cdot k_{i}+k_{2}\cdot k_{i}\right)m\left(12\dots\ell\left|12i\dots\ell\right.\right)\cr
&& 
+\dots+\left(k_{1}\cdot k_{i}+k_{2}\cdot k_{i}\dots+k_{\ell}\cdot k_{i}\right)m\left(12\dots\ell\left|12\dots\ell i\right.\right)  =0.
\end{eqnarray}
Equation~(\ref{eq:domain2sum}) then reduces to just the factor 
\begin{equation}
\alpha'^{-6}\frac{k_{1}\cdot k_{3}k_{1}\cdot k_{2}}{\left(k_{1}+k_{4}+k_{6}\right)\cdot p\left(k_{1}+k_{3}+k_{4}+k_{5}+k_{6}+k_{7}\right)\cdot p}.\label{eq:domain2result}
\end{equation}
Similarly, we can evaluate
the $\alpha'$ leading order of the integral on domain (\ref{eq:domain1}),
which gives
\begin{equation}
\alpha'^{-6}\frac{k_{1}\cdot k_{2}\left(k_{1}+k_{2}\right)\cdot k_{3}}{\left(k_{1}+k_{4}+k_{6}\right)\cdot p\left(k_{1}+k_{2}+k_{4}+k_{6}\right)\cdot p}.\label{eq:domain1result}
\end{equation}

\subsubsection{The Shapovalov form}\label{sec:shapovalov}

An alternative procedure to organise this simplification that uses the properties of the kinematic algebra
 is to write the numerator of the formula~(\ref{eq:domain2sum}) in terms of Shapovalov form 
$\left\langle \,,\,\right\rangle $, which is defined
recursively by~\cite{Fu:2022esi} 
\begin{equation}
\left\langle e_{i},e_{j}\right\rangle =\delta_{ij},\qquad\left\langle \left[e_{i},A\right],B\right\rangle =\left\langle A,\left[f_{i},B\right]\right\rangle .\label{eq:sfdef}
\end{equation}
The $e_{i}$'s here are the (positive) root vectors of the classical 
Lie algebra with root $k_{i}$, satisfying the following defining properties
\begin{equation} 
\left[e_{i},f_{j}\right] = \delta_{ij}h_{k_{i}},\qquad 
\left[h_{k_{i}},e_{j}\right] = k_{i}\cdot k_{j}e_{j},\qquad 
\left[h_{k_{i}},f_{j}\right] = -k_{i}\cdot k_{j}f_{j},\qquad 
i = 1,\dots,n.\label{eq:LAdef}
\end{equation}
The structure constants of this Lie algebra are given by scalar products of momenta. In the context of the HEFT numerators, these will be the momenta of the vertex operators of the gluons and the massive tachyons.

It was explained in~\cite{Fu:2020frx} that the vertex operators $E(k_{i})$'s form a representation of the positive root vectors, and that the negative root vectors $F(k_{i})$'s and the Cartan subalgebra $H(k_{i})$'s can be constructed accordingly so that together they form a
$q$-deformed Lie algebra. The deformation parameter
$q=e^{-i\alpha'\pi}$  {becomes unity} in the
$\alpha'\rightarrow 0$ limit,  {so that the} $q$-deformed Lie algebra reduces to the classical Lie algebra associated with the same root system. So that we may regard the generators $e_i$'s in~(\ref{eq:LAdef}) as the $\alpha'\rightarrow 0$ limit of the vertex operators $E(k_i)$ constructed in section~\ref{sec:kinematicalgebra}. 

{Alternatively, }
one can simply regard the $e_{i}$'s as formal objects, and use the algebra to organise calculations. We only use the knowledge that momentum kernels and propagator matrices can be conveniently written in terms of Shapovalov forms (for more details 
{see}
~\cite{Fu:2022esi}), to construct the numerator factors from the integration rules of section~\ref{sec:integrals}.

In terms of this language, the numerator of equation~(\ref{eq:domain2sum}) reads
\begin{equation}
\left\langle \left[e_{7}\,e_{6}\,e_{5}\,e_{4}\,e_{3}\,e_{2}\,e_{1}\right],\left[e_{2}^{*},\left[\left[e_{7}\,e_{5}\,e_{3}\right]^{*},\left[e_{6}\,e_{4}\,e_{1}\right]^{*}\right]\right]\right\rangle .\label{eq:sfexpr}
\end{equation}
We have additionally introduced the shorthand notation $\left[e_{7}\,e_{6}\,e_{5}\,e_{4}\,e_{3}\,e_{2}\,e_{1}\right]$
to denote the nested Lie-brackets $\left[e_{7}\dots\left[e_{3},\left[e_{2},e_{1}\right]\right]\right]$,
and a star  $\left[e_{6}\,e_{4}\,e_{1}\right]^{*}$ to denote the
Shapovalov dual of the nested bracket $\left[e_{6}\,e_{4}\,e_{1}\right]$
in the sense that
\begin{equation}
e_i^*=e_i,\quad
\left\langle \left[e_{6}\,e_{4}\,e_{1}\right],\left[e_{6}\,e_{4}\,e_{1}\right]^{*}\right\rangle =1,\quad\left\langle \left[e_{4}\,e_{6}\,e_{1}\right],\left[e_{6}\,e_{4}\,e_{1}\right]^{*}\right\rangle =0.
\end{equation}
In terms of Shapovalov forms, the KK-relation and the off-shell colour-kinematic
relations can be packed into the recursive relation for the dual brackets~\cite{Fu:2022esi}
\begin{equation}
\left[f_{i},\left[e_{\ell}\dots e_{i}\dots e_{2}\,e_{1}\right]^{*}\right]=\delta_{i,\ell} \, \left[e_{\ell-1}\dots e_{2}\,e_{1}\right]^{*}.\label{eq:dualrecur}
\end{equation}
Therefore, 
{we can} 
move the $e_{i}$'s in the left slot of the Shapovalov
form in 
equation~(\ref{eq:sfexpr}) to the right slot, one by one,
using the defining property of the Shapovalov form~(\ref{eq:sfdef}),
and then simplify using~(\ref{eq:dualrecur}) and~(\ref{eq:LAdef}).

Let's illustrate this by calculating the numerator of~(\ref{eq:domain2sum}) and reproducing its expression in~(\ref{eq:domain2result}).
Starting from~\eqref{eq:sfexpr}, we apply~\eqref{eq:dualrecur} to remove $e_7$, $e_6$, $e_5$ and $e_4$ to obtain
\begin{equation}
\left\langle \left[e_{3}\,e_{2}\,e_{1}\right],\left[e_{2},e_{3},e_{1}\right]\right\rangle .
\end{equation}
Using  the second commutator relation in~\eqref{eq:sfdef}, we move  $e_3$ to the right and convert it into $f_3$. Then we use the commutator $[f_{3},e_{3}]=-h_{3}$ between $f_3$ and the $e_3$ in the right bracket, to 
get the Cartan subalgebra element $h_{3}$, which  then  acts
on all the root vectors on its right, producing scalar products between the external momenta, and we obtain  
\begin{equation}
\frac{k_{3}\cdot k_{\theta(3)}}{k_{\text{all left to }3}\cdot p}\,\frac{k_{2}\cdot k_{\theta(2)}}{k_{\text{all left to }2}\cdot p}.
\end{equation}
Since all root vectors with greater labels must have been cancelled,
the result is $k_{3}\cdot k_{\theta(3)}$ where $k_{\theta(3)}$ is the
sum of all momenta with label $i$ arising from the $e_i$ that are on
the right of $e_3$. {We proceed in the} same {way} with $e_2$.

\bigskip

Note that it is straightforward to see the general pattern, especially
from the Shapovalov form perspective. A Shapovalov form such as~(\ref{eq:sfexpr})
readily translates into the rooted tree graphs of section~\ref{sec:integrals}. 
The  nested commutators in the right slot of the Shapovalov form correspond  to
the  branches rooted at the label of one of the massive tachyon, for instance at position $8$,
$$
\includegraphics[width=3cm]{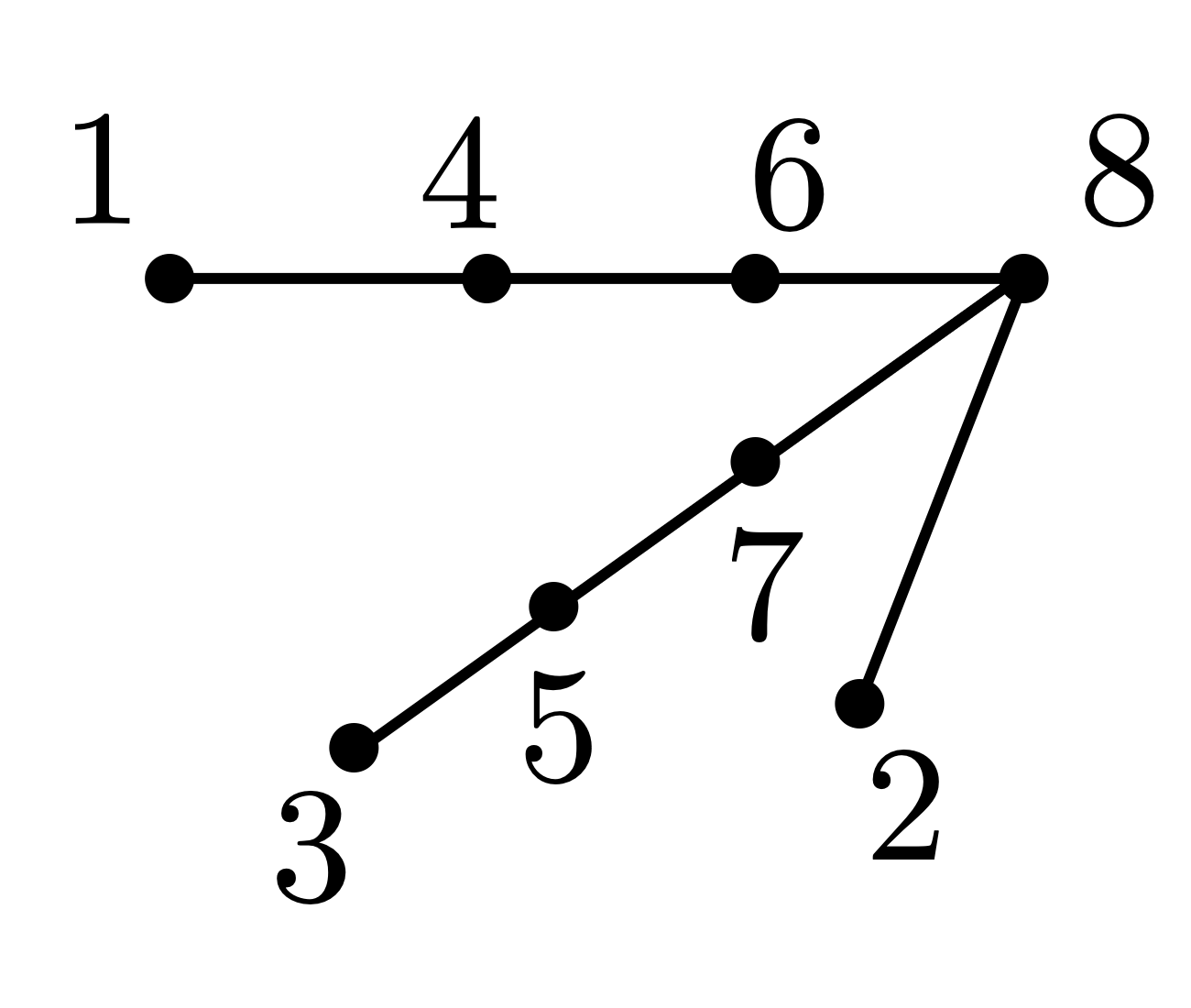}
$$

\subsection{polarisation dependence: lining up the gluons}
\label{sec:queues}

A generic tensor term obtained by applying the algorithmic procedure of section~\ref{sec:algorithm} reads 
\begin{equation}
    \prod_r  \mathcal T_r \cdot X'(t_r),
\end{equation}
where the tensors $ \mathcal T_r$ are the product of polarisation tensors
and field-strength. 
{Here}
we do not write the poles, the Koba-Nielsen
factors and the plane-wave exponential factors, because we are
focusing on the tensorial pieces from the contractions of the vertex
operators.
As a consequence of the commutator relation presented before and the IBP procedure used in equation~\eqref{e:V1vecV2vec}, the right-most factor of the tensors $\mathcal T_r$ is always the field-strength $F_r$ of the gluon with label $r$.

When taking the commutator with a new gluon {vertex operator}, the newly introduced polarisation tensor $\epsilon_{i}$ can take one of the following three actions:
\begin{itemize}
    \item[(i)] Becoming the last entry in the tensor, so that we get $\alpha' \mathcal T_r\cdot F_i \cdot X'(t_i)$,
    \item[(ii)]  Starting a new product term of its own $\epsilon_{i}\cdot X'$,
    \item[(iii)]   Contracting with one of the existing exponentials, which contributes a factor $\alpha' \epsilon_{i}\cdot k_{j}/(t_{i}-t_{j})$.
\end{itemize}
All other alternatives lead to sub-leading contributions in $\alpha'$ or eventually to vanishing integrals because the graphs do not satisfy the ascending order condition.

\medskip

The expressions derived in the previous section are expressed in terms of gluon polarisation tensors, and are not in explicit gauge invariant form. The  numerator factors in~\eqref{e:Nstring} are by definition gauge invariant  and can only be expressed in terms of field-strengths. We have two ways of doing this. One is to choose the reference momentum for each gluon vertex operator so that the polarisation tensor is given by $\bar\epsilon_i^\mu=(p\cdot F_i)^\mu/(p\cdot k_i)$, or systematically use integration-by-parts identities to pack the polarisation tensors into field-strengths. In both cases,  the total derivatives  
do not contribute to the result because when integrating over the contour $\mathcal C$ 
{the boundary term collides with} 
the massive (tachyonic) particle fixed at position $1$.
{Since the resulting Koba-Nielsen factor contains a factor $(t_i -
  1)^{\alpha' k_{i}\cdot k_{n+1}}$ evaluated at $t_{i}\rightarrow 1$,
  this gives a vanishing contribution.}

Both ways of writing the expression in terms of field-strengths bring a lot of spurious poles. These can be reduced by combining the choice of the reference momentum and the IBP depending on the position of the vertex operator in the integration string. 
This approach is presented in the next section.


\section{Manifestly gauge invariant approach} \label{sec:manifest}

In this section, we  describe how the rules are affected if all polarisation of the gluons are gauge fixed to be of the form $\bar\epsilon_i^\mu= (p\cdot
F)^\mu/(p\cdot k_i)$.

In this approach, the $\alpha'$ leading order contributions are (again) labelled by ascending tree graphs, but every branch of a tree is restricted to contain at least two labels. For an $n$-gluon amplitude, each term corresponds to an ascending tree with its root labelled by $n+1$, and the rules for computing the corresponding $\alpha'$ leading order contribution are as follows. For each rooted tree, we first manually assign an order to all its branches, with the branch containing the label $1$ being fixed as the first. In each of the orders, we denote the labels on the $i$-th branch by $\tau_{i}$. Additionally, we denote the labels sitting on bra{n}ch $\tau_{i}$ by $\left(a_{i}\left(1\right)\dots a_{i}\left(s_{i}\right)\right)$, where $a_{i}(s_{i})$ is the label adjacent to the root and $s_{i}$
is the length of this branch. We shall further refer to all the remaining
edges that connect to different branches (but not to the root) as leaves,
and we denote them by $\left[c_{i}\left(1\right)\right]_{b_{i}\left(1\right)}\left[c_{i}\left(2\right)\right]_{b_{i}\left(2\right)}\dots\left[c_{i}\left(l_{i}\right)\right]_{b_{i}\left(l_{i}\right)},\ c_{i}\left(\ell\right)>b_{i}\left(\ell\right)$,
where $c_{i}\left(\ell\right)$ and $b_{i}\left(\ell\right)$ are
the labels corresponding to the two ends of the leaves, $b_{i}\left(\ell\right)$
represents the location where the leaf connects to the branch
$\tau_{i}$, and $l_{i}$ is the total number of the leaves on this
branch. So, generically, any branch $\tau_{i}$ can be expressed
as $\left(a_{i}\left(1\right)\dots a_{i}\left(l_{i}\right)\right)\left[c_{i}\left(1\right)\right]_{b_{i}\left(1\right)}\left[c_{i}\left(2\right)\right]_{b_{i}\left(2\right)}\dots\left[c_{i}\left(\ell\right)\right]_{b_{i}\left(\ell\right)}$.
For such a generic tree graph, we write down its corresponding leading
order contribution in $\alpha'$ as follows
\begin{itemize}
\item Each branch $\tau_{i}$ containing the labels $\left(a_{i}\left(1\right)\dots a_{i}\left(s_{i}\right)\right)$
contributes a factor  of
\begin{equation}
\alpha'^{s_{i}} \,\frac{p\cdot F_{a_{i}\left(1\right)}\cdot\ldots \cdot F_{a_{i}\left(s_{i}\right)}\cdot p}{k_{s_{i}\left(1\right)}\cdot p},
\end{equation}
\item Each leaf $\left[c_{i}\left(\ell\right)\right]_{b_{i}\left(\ell\right)}$
contributes a factor 
\begin{equation}
\alpha'\,\frac{p\cdot F_{c_{i}\left(\ell\right)}\cdot k_{b_{i}\left(\ell\right)}}{k_{c_{i}\left(\ell\right)}\cdot p},
\end{equation}
\end{itemize}
If $i>1$, then a branch $\tau_{i}=\left(a_{i}\left(1\right)\dots a_{i}\left(s_{i}\right)\right)\left[c_{i}\left(1\right)\right]_{b_{i}\left(1\right)}\left[c_{i}\left(2\right)\right]_{b_{i}\left(2\right)}\dots\left[c_{i}\left(l_{i}\right)\right]_{b_{i}\left(l_{i}\right)}$
contributes additionally a factor
\begin{equation}
\frac{k_{a_{i}\left(1\right)}\cdot k_{\theta\left(a_{i}\left(1\right)\right)}}{\sum_{j\in\tau_{1}\cup\tau_{2}\dots\cup\tau_{i-1}}k_{j}\cdot p},
\end{equation}
The contribution of the whole tree is given by the product of all the factors listed above and then summed over all possible permutations of the branches. 

For example, in the $7$-gluon amplitude, we have the following ascending
tree rooted at label $8$.
$$
\includegraphics[width=2.5cm]{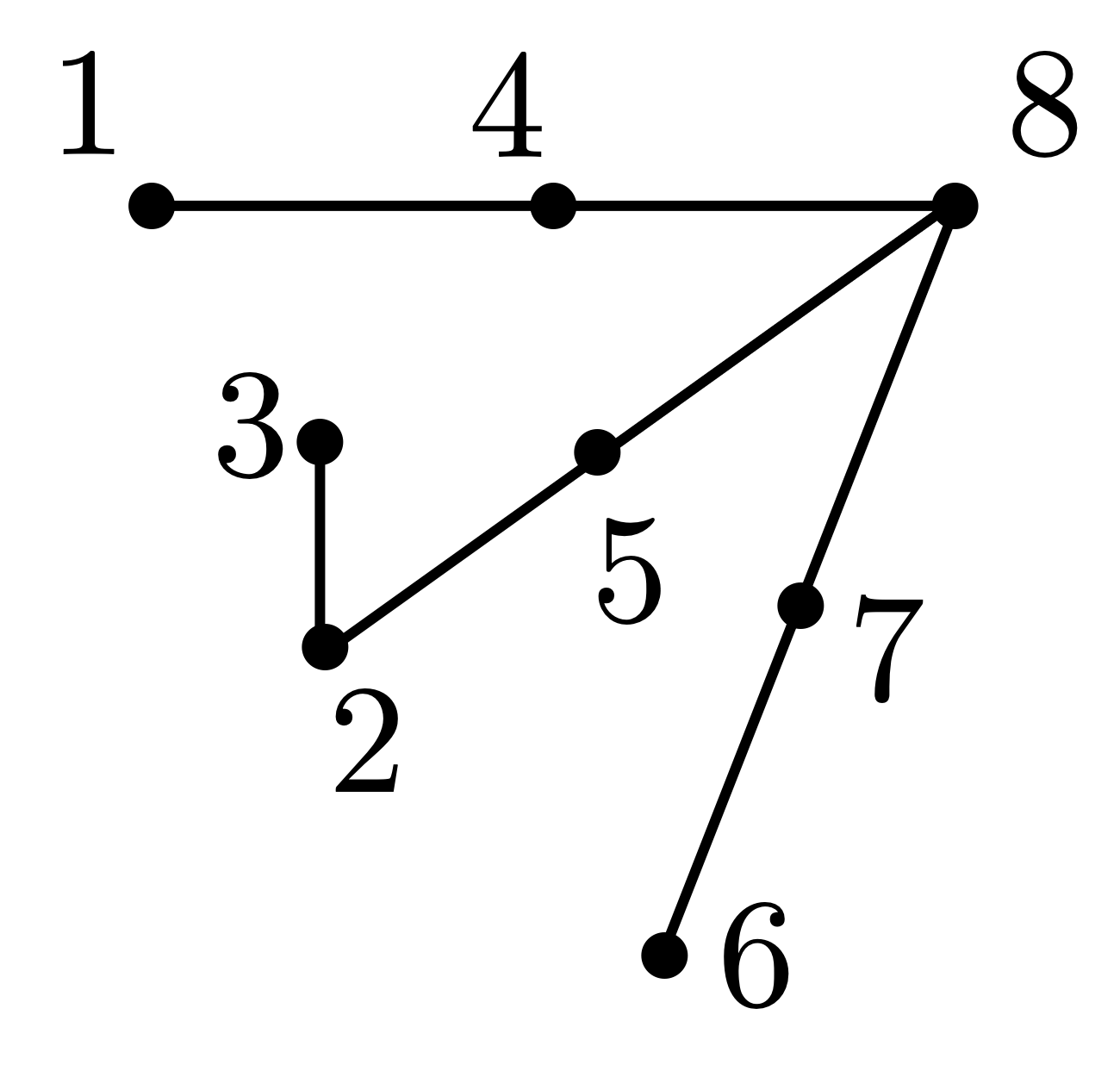}
$$
The tree branches associated with this graph are
\begin{equation}
\left(14\right),\quad\left(258\right)\left[3\right]_{2},\quad\left(67\right).
\end{equation}
We need to consider two possible permutations of the branches 
\begin{equation}
\tau_{1}=\left(14\right),\quad\tau_{2}=\left(25\right)\left[3\right]_{2},\quad\tau_{3}=\left(67\right),
\end{equation}
and
\begin{equation}
\tau_{1}=\left(14\right),\quad\tau_{2}=\left(67\right),\quad\tau_{3}=\left(25\right)\left[3\right]_{2}.
\end{equation}
According to the rules described above, the two permutations correspond
to the following product of factors, respectively
\begin{multline}
\alpha'^{7} \, \frac{p\cdot F_{1}\cdot F_{4}\cdot p}{k_{1}\cdot p}  \frac{k_{1}\cdot k_{2}}{k_{1}\cdot p+k_{4}\cdot p}\frac{p\cdot F_{2}\cdot F_{5}\cdot p}{k_{2}\cdot p}\frac{p\cdot F_{3}\cdot k_{2}}{k_{3}\cdot p}\cr
  \times\frac{k_{1}\cdot k_{6}+k_{2}\cdot k_{6}+k_{3}\cdot k_{6}+k_{4}\cdot k_{6}+k_{5}\cdot k_{6}}{k_{1}\cdot p+k_{2}\cdot p+k_{3}\cdot p+k_{4}\cdot p+k_{5}\cdot p}\frac{p\cdot F_{6}\cdot F_{7}\cdot p}{k_{6}\cdot p},
\end{multline}
and 
\begin{multline}
\alpha'^{7} \, \frac{p\cdot F_{1}\cdot F_{4}\cdot p}{k_{1}\cdot p} \frac{k_{1}\cdot k_{6}+k_{4}\cdot k_{6}}{k_{1}\cdot p+k_{4}\cdot p}\frac{p\cdot F_{6}\cdot F_{7}\cdot p}{k_{6}\cdot p}\cr
 \times \frac{k_{1}\cdot k_{2}}{k_{1}\cdot p+k_{4}\cdot p+k_{6}\cdot p+k_{7}\cdot p}\frac{p\cdot F_{2}\cdot F_{5}\cdot p}{k_{2}\cdot p}\frac{p\cdot F_{3}\cdot k_{2}}{k_{3}\cdot p}.
\end{multline}


\section{Conclusion}\label{sec:conclusion}

We have given a derivation of the HEFT numerators from the $\alpha'\to0$ 
limit of the kinematic vertex operator algebra introduced in~\cite{Fu:2018hpu}.
The construction of HEFT  numerators for high multiplicity is a non-trivial task.  This is partly due to the non-uniqueness of the numerator,  
{which} 
depends on the choice of gauge, and the large number of terms involved. 
Because the numerator factors arise from multiple nested commutators
of $n$ gluon vertex operators, they are organised in terms of a rooted
tree graph endowed with an ordering, which facilitates the
enumeration of the various contributions.  We have written a
\texttt{Mathematica} code for producing these numerator
factors. {The code and examples are} available at this repository~\href{https://github.com/Yi-hongWang/Stringy-Numerator}{Stringy-Numerator}.

\medskip

The field theory fusion rules of~\cite{Brandhuber:2021kpo,Brandhuber:2021bsf} 
{were} 
derived from the field theory expression of the numerator factors which is the $\alpha'\to0$ limit of~\eqref{e:Nstring}. Consequently, the kinematic algebra detailed in section~\ref{sec:heft-algebra} gives a string theory origin of their construction. 
Although, by definition,  the numerator factors must  be the same
gauge invariant object{s},  they can take algebraically different
forms. For example, the two constructions can have different sets of
spurious poles. Some of these poles can be removed by  partial
fractioning and using   the following identity  involving the    fieldstrength of gluon $i$ 
\begin{equation}
A\cdot k_i \,\left(P\cdot F_i\over P\cdot k_i\right)\cdot B-A\cdot \left(P\cdot F_i\over P\cdot k_i\right) k_i \cdot B=A\cdot F_i\cdot B,
\end{equation}
which is derived by using the gauge
{choice in eq.}~\eqref{e:barepsilon}  for the
gluon $i$ with 
reference momentum $P$  {not parallel to} $k_i$.

\medskip
We have presented 
the construction of kinematic numerator factors from the vertex
operator algebra  for the scalar line emissions case in
detail. {We have}   explained how this extends to the
fermionic line case, {and showed how to reproduce}  the results of~\cite{Bjerrum-Bohr:2024fbt}. The extension
to the {multi-gluon} emission from {a} massive tensor particles can be done along the same lines. 
The nature of the massive external states only affects the final steps 
{when we evaluate} 
the expectation value.

Also, since the construction of the kinematic algebra is done in string theory, we can preserve the $\alpha'$ corrections to the numerator factors~\cite{Chen:2024gkj}, which could be useful for analysing the effects of higher derivative corrections from string theory on gravitational wave observables.


\begin{acknowledgments}
We thank Emil Bjerrum-Bohr for helpful correspondance.
 PV would like to thank the LAPTh for hospitality when this work was started. CF and YW would like to express gratitude to Jianxin Lu and Yang Zhang for their invitation to the Peng Huanwu Center for Fundamental Theory, and to Yi-Jian Du for the invitation to Wuhan University. CF and YW also
acknowledge Gang Chen and Roman Lee for insightful discussions. 
CF is supported by the High-level Talent 
Research Start-up Project Funding of Henan Academy of Sciences
(Project No. 241819244). YW is supported by China National Natural Science Funds for Distinguished Young Scholar (Grant No. 12105062) and Agence Nationale de la Recherche (ANR), project ANR-22-CE31-0017. PV
was supported in part by French National Agency for Research
grant ``Observables'' (ANR-24-CE31-7996).
\end{acknowledgments}


\appendix
\section{Comparison the HEFT numerators from fusion rules}
\label{sec:matchQMUL}

In this appendix, we enumerate all the terms that appear at three- and
four-gluon integrands and the final expressions of string HEFT numerators. 

The algorithmic procedure presented in the main text gives a simple bookkeeping for enumerating all the contributions. We explain how this reproduces the results of~\cite{Brandhuber:2021kpo,Brandhuber:2021bsf}.

In the following discussion, we assume the algorithm explained in sections \ref{sec:algorithm} to \ref{sec:queues},
but do not shift polarisations into manifestly gauge invariant form as in section \ref{sec:manifest}.
The shorthand notations we introduced there
also change accordingly:
The parenthesis $\left(a_{i}\left(1\right)\dots a_{i}\left(s_{i}\right)\right)$ now stands for the product $\left( \epsilon_{a_{i}\left(1\right)}\cdot\ldots \cdot F_{a_{i}\left(s_{i}\right)}\cdot p \right)$, with the 
first label associated with a polarisation instead of
 field strength.
The square bracket $\left[c_{i}\left(\ell\right)\right]_{b_{i}\left(\ell\right)}$
now represents a factor 
$\epsilon_{c_{i}\left(\ell\right)}\cdot k_{b_{i}\left(\ell\right)}$.
Besides polarisation and momentum dependence, a term that appears in the algorithm  also carries a world-sheet integral which, as was explained in the section
\ref{sec:integrals}, can be expressed as a tree graph.
The branches of a tree graph are no longer restricted to contain at least two labels in the case where we do not shift polarisations. The ascending tree condition
still applies because the world-sheet integrals are carried out in the same way. For simplicity, let us introduce additionally the following notation for the Parke-Taylor factor
\begin{equation}
    \frac{1}{{\rm PT}(\sigma_{1} \sigma_{2} \dots \sigma_{r})} :=\frac{1}{
    (t_{\sigma_{1}}-t_{\sigma_{2}})(t_{\sigma_{2}}-t_{\sigma_{3}}) \dots (t_{\sigma_{r-1}}-t_{\sigma_{r}})
    }
\end{equation}

\medskip
\noindent $\bullet$ {\bf The three-gluon case:}
Bearing the ascending condition in mind,
we see there are only six terms that 
carry non-vanishing integrands in
three-gluon emission problem
\begin{multline}
  \frac{\left(13\right)\left[2\right]_{1}}{\mathrm{PT}\left(134\right)\mathrm{PT}\left(12\right)}-\frac{\left(13\right)\left(2\right)}{\mathrm{PT}\left(134\right)\mathrm{PT}\left(24\right)}+\frac{\left(123\right)}{\mathrm{PT}\left(1234\right)}
 +
 \frac{\left(12\right)\left[3\right]_{1}}{\mathrm{PT}\left(124\right)\mathrm{PT}\left(13\right)}\cr
 +\frac{\left(12\right)\left[3\right]_{2}}{\mathrm{PT}\left(124\right)\mathrm{PT}\left(23\right)}
 -\frac{\left(12\right)\left(3\right)}{\mathrm{PT}\left(134\right)\mathrm{PT}\left(24\right)}.
 \label{eq:3gluon-ex-1}
\end{multline}
The result of world-sheet integrals can also be expressed using a compact notation. Recall from section~\ref{sec:integrals}, that these integrals can be characterised by tree graphs, whose $\alpha'$ leading order contributions can in turn be read off from graphs, using formula (\ref{eq:integral-rules}).
In light of this, let us introduce square brackets 
$\langle a_{i}\left(1\right)\dots a_{i}\left(s_{i}\right)\rangle_{i}$ to denote an
$i$-th tree branch. A tree graph containing $r$ branches is therefore represented as $\langle a_{1}\left(1\right)\dots a_{1}\left(s_{1}\right)\rangle_{1} \, \dots 
\langle a_{i}\left(1\right)\dots a_{r}\left(s_{r}\right)\rangle_{r}$.
We make a distinction here between the parenthesis
associated with polarisation dependence and the 
angle brackets associated with the pole structure of 
world-sheet integrals because the later may change
after we incorporate the additional poles introduced by
the leaves $\left[c_{i}\left(\ell\right)\right]_{b_{i}\left(\ell\right)}$ and simplify using partial fractions,
even though the parenthesis and the angle brackets
should coincide prior to this procedure.

When the world-sheet integrals are expressed in terms of angle brackets, the above equation (\ref{eq:3gluon-ex-1}) becomes the following
\begin{multline}
     \left\langle 123\right\rangle _{1}\left(13\right)\left[2\right]_{1}-\left\langle 13\right\rangle _{1}\left\langle 2\right\rangle _{2}\left(13\right)\left(2\right)+\left\langle 123\right\rangle _{1}\left(123\right)
  + \left\langle 123\right\rangle _{1}\left(12\right)\left[3\right]_{1} 
 +\left\langle 123\right\rangle _{1}\left(12\right)\left[3\right]_{2}\cr
 -\left\langle 12\right\rangle _{1}\left\langle 3\right\rangle _{2}\left(12\right)\left(3\right) =  \langle T_{\left(123\right)}-T_{\left(12\right),\left(3\right)}-T_{\left(13\right),\left(2\right)} \rangle.
\end{multline}
Substituting angle brackets using formula using formula (\ref{eq:integral-rules}) and we have an agreement with the fusion rules result of~\cite{Brandhuber:2021kpo,Brandhuber:2021bsf}
\begin{equation}
    \langle T_{(\tau_{1}),(\tau_{2}),\dots ,(\tau_{r})}\rangle 
    :=
    \frac{ p\cdot F_{\tau_{1}}\cdot V_{\theta(\tau_{2})} 
    \cdot F_{\tau_{2}} \cdots 
    F_{\tau_{r}} \cdot p
    }{(n-2)\,k_{1}\cdot p \, k_{\tau_{1}}\cdot p 
    \, \cdots k_{\tau_{1}\tau_{2}\dots\tau_{r}}\cdot p}.
\end{equation}

\medskip
\noindent $\bullet$ {\bf The four-gluon case:} 
For the  four-gluon case, the leading contribution to integrand consists of the following
$39$ terms:
\begin{eqnarray}
    && \frac{\left(123\right)\left(4\right)}{\mathrm{PT}\left(1235\right)\mathrm{PT}\left(45\right)},\frac{\left(124\right)\left(3\right)}{\mathrm{PT}\left(1245\right)\mathrm{PT}\left(35\right)},\frac{\left(134\right)\left(2\right)}{\mathrm{PT}\left(1345\right)\mathrm{PT}\left(25\right)},
    \frac{\left(12\right)\left(34\right)}{\mathrm{PT}\left(125\right)\mathrm{PT}\left(345\right)},
    \cr
    && 
    \frac{\left(13\right)\left(24\right)}{\mathrm{PT}\left(135\right)\mathrm{PT}\left(245\right)},\frac{\left(14\right)\left(23\right)}{\mathrm{PT}\left(145\right)\mathrm{PT}\left(235\right)},
    \frac{\left(12\right)\left(3\right)\left(4\right)}{\mathrm{PT}\left(125\right)\mathrm{PT}\left(35\right)\mathrm{PT}\left(45\right)},
     \cr
   &&
   \frac{\left(13\right)\left(2\right)\left(4\right)}{\mathrm{PT}\left(135\right)\mathrm{PT}\left(25\right)\mathrm{PT}\left(45\right)},
  \frac{\left(14\right)\left(2\right)\left(3\right)}{\mathrm{PT}\left(145\right)\mathrm{PT}\left(25\right)\mathrm{PT}\left(35\right)},\frac{\left(1234\right)}{\mathrm{PT}\left(12345\right)}, 
   \cr
   &&
   \frac{\left(123\right)\left[4\right]_{b_{1}}}{\mathrm{PT}\left(1235\right)\mathrm{PT}\left(b_{1}4\right)},\frac{\left(124\right)\left[3\right]_{a_{1}}}{\mathrm{PT}\left(1245\right)\mathrm{PT}\left(a_{1}3\right)},\frac{\left(134\right)\left[2\right]_{1}}{\mathrm{PT}\left(1345\right)\mathrm{PT}\left(12\right)},
 \frac{\left(12\right)\left[3\right]_{a_{2}}\left(4\right)}{\mathrm{PT}\left(125\right)\mathrm{PT}\left(a_{2}3\right)\mathrm{PT}\left(45\right)},
 \cr
   &&  \frac{\left(12\right)\left(3\right)\left[4\right]_{b_{2}}}{\mathrm{PT}\left(125\right)\mathrm{PT}\left(35\right)\mathrm{PT}\left(b_{2}4\right)},
   \frac{\left(12\right)\left[3\right]_{a_{3}}\left[4\right]_{b_{3}}}{\mathrm{PT}\left(125\right)\mathrm{PT}\left(a_{3}3\right)\mathrm{PT}\left(b_{3}4\right)},
  \frac{\left(13\right)\left[2\right]_{1}\left(4\right)}{\mathrm{PT}\left(135\right)\mathrm{PT}\left(12\right)\mathrm{PT}\left(45\right)},
   \cr
    && \frac{\left(13\right)\left(2\right)\left[4\right]_{b_{4}}}{\mathrm{PT}\left(135\right)\mathrm{PT}\left(35\right)\mathrm{PT}\left(45\right)},\frac{\left(13\right)\left[2\right]_{1}\left[4\right]_{b_{5}}}{\mathrm{PT}\left(135\right)\mathrm{PT}\left(12\right)\mathrm{PT}\left(b_{5}4\right)},
    \frac{\left(14\right)\left(2\right)\left[3\right]_{a_{4}}}{\mathrm{PT}\left(145\right)\mathrm{PT}\left(25\right)\mathrm{PT}\left(a_{4}3\right)},
     \cr
    && \frac{\left(14\right)\left[2\right]_{1}\left(3\right)}{\mathrm{PT}\left(145\right)\mathrm{PT}\left(12\right)\mathrm{PT}\left(35\right)},
    \frac{\left(14\right)\left[2\right]_{1}\left[3\right]_{a_{5}}}{\mathrm{PT}\left(145\right)\mathrm{PT}\left(12\right)\mathrm{PT}\left(a_{5}3\right)},\quad a_{i}\in\left\{ 1,2\right\} ,\ b_{i}\in\left\{ 1,2,3\right\} 
    \label{eq:4gluon-ex-1}
\end{eqnarray}
Evaluating these terms using the set of integration rules of section~\ref{sec:integrals}
leads  again  to identical results to those in~\cite{Brandhuber:2021bsf}, which can be
checked by verifying that the coefficients of each contraction factor
calculated from equation{s} (5) and (9) in~\cite{Brandhuber:2021bsf} 
add up to their corresponding
integration rule factor~(\ref{eq:integral-rules}).

\medskip
For example, in equation (\ref{eq:4gluon-ex-1}), the term that carries momentum and polarisation dependence
$\left(14\right)\left[2\right]_{1}\left(3\right)=(\epsilon_{1}\cdot F_{4}\cdot p)(k_{1}\cdot\epsilon_{2})(\epsilon_{3}\cdot p)$
evaluates to 
\begin{equation}
\frac{\left(14\right)\left[2\right]_{1}\left(3\right)}{\mathrm{PT}\left(145\right)\mathrm{PT}\left(12\right)\mathrm{PT}\left(35\right)}  \rightarrow\left\langle 124\right\rangle _{1}\left\langle 3\right\rangle _{2}\left(14\right)\left[2\right]_{1}\left(3\right)
  =\frac{\left(k_{1}+k_{2}\right)\cdot k_{3}}{\left(k_{1}+k_{2}+k_{4}\right)\cdot p}\left(14\right)\left(3\right)\left[2\right]_{1}.
\end{equation}
On the other hand, there are three terms produced by the fusion rules
that carry the same contraction factor
$\left(14\right)\left[2\right]_{1}\left(3\right)$ {given by}
\begin{equation}
T_{\left(14\right),\left(2\right),\left(3\right)},\ T_{\left(14\right),\left(3\right),\left(2\right)},\ -T_{\left(14\right),\left(23\right)}.
\end{equation}
The coefficients of $\left(14\right)\left[2\right]_{1}\left(3\right)$
of these three terms are
\begin{equation}
 -\frac{k_{2}\cdot p\left(k_{1}+k_{2}\right)\cdot k_{3}}{\left(k_{1}+k_{4}\right)\cdot p\left(k_{1}+k_{2}+k_{4}\right)\cdot p},\ 
  -\frac{k_{2}\cdot p k_{1}\cdot k_{3}}{\left(k_{1}+k_{4}\right)\cdot p\left(k_{1}+k_{3}+k_{4}\right)\cdot p},\ \frac{k_{2}\cdot k_{3}}{\left(k_{1}+k_{4}\right)\cdot p}\label{eq:6pttcoeff}
\end{equation}
respectively. It is straightforward to check that the sum of {the} coefficients
in~(\ref{eq:6pttcoeff}) equals $\left\langle 124\right\rangle _{1}\left\langle 3\right\rangle _{2}$,
up to {the} mass-shell condition of the two tachyons.

\medskip

We have performed the same consistency check term by term up to the eight
gluons case
using the \texttt{Mathematica} code  {available} here~\href{https://github.com/Yi-hongWang/Stringy-Numerator}{GitHub}.


\end{document}